\documentclass[journal]{IEEEtran}

\usepackage[utf8]{inputenc}
\usepackage{enumitem}
\usepackage{dsfont}
\usepackage[font=small]{caption}
\usepackage{color}  

\usepackage[nolist]{acronym}
\begin{acronym}
    \acro{MIR}{Music Information Retrieval}
    \acro{STFT}{Short Time Fourier Transform}
    \acro{CNN}{Convolutional Neural Network}
    \acrodefplural{CNN}{Convolutional Neural Networks}
    \acro{DNN}{Deep Neural Network}
    \acrodefplural{DNN}{Deep Neural Networks}
    \acro{CQT}{Constant-Q transform}
    \acro{MAP}{Mean Average Precision}
    \acro{DTW}{Dynamic Time Warping}
\end{acronym}

\acrodef{dtw}[DTW]{Dynamic Time Warping}
\acrodef{ssl}[SSL]{Self Supervised Learning}
\acrodef{vm}[V/M]{Video/Music}
\acrodef{v2m}[V2M]{Video-to-Music}
\acrodef{av}[AV]{Audio/Video}
\acrodef{cca}[CCA]{Canonical Correlation Analysis}

\usepackage[noadjust]{cite} % to automatically sort and compress citations.

\usepackage{graphicx} % For graphics, photos, etc

\usepackage{amsmath}
\usepackage{cases}

\usepackage[ruled,vlined]{algorithm2e}

\usepackage{array}

\usepackage[caption=false,font=footnotesize]{subfig}

\usepackage{multirow}
\usepackage{multicol}
\usepackage{tabularx}
\usepackage{caption}

\usepackage{hyperref}  
\usepackage{url}

\begin{document}

% Title.
% ------
\title{Video-to-Music Recommendation using \\ Temporal Alignment of Segments}

% Author names.
% -------------
\author{Laure Prétet,~\IEEEmembership{Graduate~Student~Member,~IEEE,}
        Gaël Richard,~\IEEEmembership{Fellow,~IEEE,}\\
        Clément Souchier,
        and~Geoffroy Peeters,~\IEEEmembership{Member,~IEEE}%
        }

% The paper headers
\markboth{IEEE Transactions on Multimedia,~Vol.~N, No.~N, Month~Year}%
{Prétet \MakeLowercase{\textit{et al.}}: Video-to-Music Recommendation using Temporal Alignment of Segments}
% The only time the second header will appear is for the odd numbered pages after the title page when using the twoside option.
% *** Note that you probably will NOT want to include the author's *** name in the headers of peer review papers. 
% You can use \ifCLASSOPTIONpeerreview for conditional compilation here if you desire.

% make the title area
\maketitle

%%%%%%%%%%%%%%%%%%%%%%%%%%%%%%%%%%%%%%%%%%%%%%%%%%%%%%%
%%%%%%%%%%%%% ABSTRACT AND KEYWORDS %%%%%%%%%%%%%%%%%%%
%%%%%%%%%%%%%%%%%%%%%%%%%%%%%%%%%%%%%%%%%%%%%%%%%%%%%%%

\begin{abstract}

We study cross-modal recommendation of music tracks to be used as soundtracks for videos.
This problem is known as the music supervision task.
We build on a self-supervised system that learns a content association between music and video.
In addition to the adequacy of content, adequacy of structure is crucial in music supervision to obtain relevant recommendations.
We propose a novel approach to significantly improve the system's performance using structure-aware recommendation.
The core idea is to consider not only the full audio-video clips, but rather shorter segments for training and inference. 
We find that using semantic segments and ranking the tracks according to sequence alignment costs significantly improves the results.
We investigate the impact of different ranking metrics and segmentation methods.

\end{abstract}


\IEEEpeerreviewmaketitle

%%%%%%%%%%%%%%%%%%%%%%%%%%%%%%%%%%%%%%%%%%%%%%%%%%%%%%%
%%%%%%%%%%%%%%%%%%% BODY %%%%%%%%%%%%%%%%%%%%%%%%%%%%%%
%%%%%%%%%%%%%%%%%%%%%%%%%%%%%%%%%%%%%%%%%%%%%%%%%%%%%%%

\section{Introduction}
\label{sec:intro}

% Introducing the global topic: music-video recommendation
\IEEEPARstart{I}{n} our everyday life, co-occurrence of events across our senses allows us to efficiently acquire a large amount of knowledge without the need of labels or ground truth annotations. 
For example, we are able to associate engine sounds to cars, musical sounds to specific instruments, a sour taste to the picture of a lemon, and so on.

If some of these cross-modal associations are straightforward (e.g. the sound of a guitar with an image of guitar), some others are more subtle and may be dependant on our personal experience or taste. The literature of such cross-modal associations is vast and has been studied in the context of multiple applications ranging from cross-modal retrieval (e.g. video retrieval by textual description \cite{Gabeur2020Multi-modalRetrieval, Ma2020VLANet:Retrieval}), multimodal classification (e.g. speaker diarization \cite{Noulas2012MultimodalDiarization}), scene analysis \cite{Owens2018Audio-VisualFeatures}, cross-modal source separation \cite{Zhao2018ThePixels} or even cross-modality translation (e.g. translating music into paintings \cite{Anger2021TraduireKandinsky}).

In this work, we are interested in cross modal \ac{vm} recommendation  and more specifically in the \textit{music supervision} use case.
\ac{vm} is a vast domain which has applications in automatic video editing \cite{Shah2014ADVISORRankings}, automatic MTV generation \cite{Liao2009MiningMTV}, and general music recommendation \cite{Zeng2018Audio-VisualCCA}.

At the heart of a fast growing industry, music supervision is the task of finding
the most suitable music from a vast catalog to serve as a soundtrack for a video.
Such videos can be ads, movies, TV series~\cite{Inskip2008MusicRetrieval}, user-generated videos on video streaming services (such as TikTok) or video games.

This cross-modal retrieval task relies on specialists who use their musicological expertise and their catalog-knowledge base to make the best suggestions.
Today, the field still requires a large amount of manual work, either to annotate each track or to listen to as many tracks as possible to find the best match.

Designing an automatic music supervision system is however very challenging because it needs to integrate high level semantics of the music and video tracks. Their structural content, the invoked emotions, the location of their "climax" are some examples of these important high-level concepts for music supervision.

% Scientific scope
There is already a very rich literature on the general problem of \ac{vm}. We further discuss, in Section \ref{sec:soa}, these related works which can be grouped in three main categories: \acl{vm} matching by semantic association, by sequence comparison or by leveraging music structure.

% Motivations
Today, state-of-the-art approaches in \ac{vm} recommendation rely on the \ac{ssl} paradigm, which is inspired by the day-to-day human experience.
For example, Hong et al. \cite{Hong2018CBVMR:Constraint} proposed the VM-Net, a neural network for \acl{vm} recommendation trained using \ac{ssl}.
The VM-Net is able to learn a content association between video and music from a large amount of unlabelled music video clips. 
Compared to other similar works, this system has interesting specificities such as an extended triplet loss \cite{Weinberger2009DistanceClassification, Schroff2015FaceNet:Clustering} for video and music consistency and a lightweight architecture thanks to the use of high-level features as inputs.

However, this approach suffers from two main limitations. First, as shown in our previous work \cite{Pretet2021DesignRecommendation}, the performances of the VM-Net are limited by the original choice of handcrafted 
audio features.
Second, and more importantly, the whole duration of the video and of the music are summed up as timeless embedding vectors, i.e. the original method does not allow for representing variations of content over time.
This is a very strong simplification of the musical content that is in contradiction with the way music video clips and music tracks are constructed (as a sequence of shots and verse-chorus segments). In fact, our recent study on music video clips \cite{Pretet2021IsStudy} confirmed the intuition that a high-performing \ac{vm} recommendation system should take into consideration the structure of the music tracks and query videos.
This raises several questions:
How can we define "structure" in musical videos?
Which temporal scale should we consider to model the temporal evolution?
How can we train the network at shorter time scales?
What is the best way to evaluate it?

\begin{figure*}
    \centering
    \includegraphics[width=\textwidth]{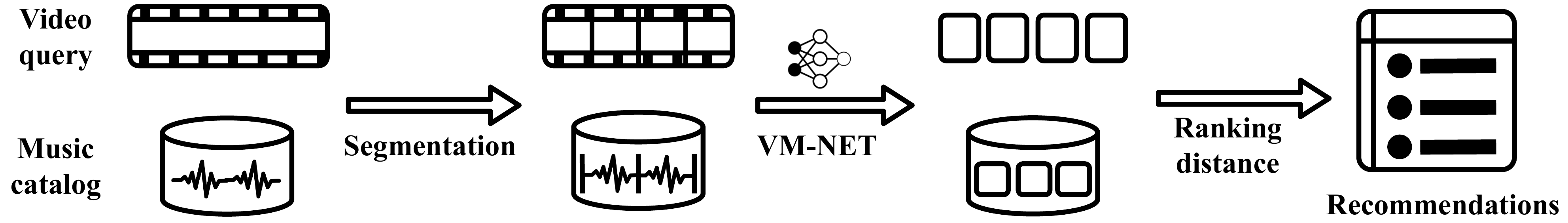}
    \caption{General overview of the studied Seg-VM-Net recommendation system (inference perspective).}
    \label{fig:system}
\end{figure*}

The main contributions of our work can be summarized as follows: 
\begin{itemize}
    \item We propose a novel \ac{v2m} system which operates on semantically consistent segments from the video and music to better exploit structural content of both modalities (see Figure~\ref{fig:system} for an overview of our Seg-VM-net approach.) Unlike previous approaches, our method allows to recommend music beyond short snippets while preventing that full songs are summarized into a unique embedding vector.
    \item Our Seg-VM-net model integrates a multimodal neural network allowing a single representation per segment, and a dynamic time warping alignment cost between two sequences of segments; 
    \item We provide an extensive study of semantic segmentation methods and segment aggregation functions suitable for audio-video music clips;
    \item We conduct an extended experimental study which shows that our method significantly improves performance and robustness compared to previous clip-level approaches.
\end{itemize}

To our knowledge, this is the first Video-to-Music recommendation system that combines a self-supervised multimodal embedding space dedicated to \acl{vm} recommendation, a dynamic time warping alignment cost between two sequences of segments and a semantic definition of segments, capable of handling units of variable durations.

Note that we assume here that an efficient music recommendation system should fulfill the following constraints: querying should be fast enough to scale to large catalogs, and the video query cannot be edited while the music tracks can (but only to a certain extent, typically by slicing them into meaningful segments).

\textbf{Paper organization.} 
In Section~\ref{sec:soa}, we review works related to ours. 
In Section~\ref{sec:baseline}, we describe the VM-Net system, which we consider as our baseline for \ac{v2m} recommendation.
In Section~\ref{sec:method}, we then describe our modifications to the VM-Net system that allow its training and inference at the segment-level (rather than at the clip-level).
We propose various segmentation paradigms, discuss triplet mining in the case of segments, and propose various segment aggregation and alignment algorithms.
In Section~\ref{sec:eval}, we propose an experimental protocol which allows evaluating \ac{v2m} systems in a more realistic scenario.
In Section~\ref{sec:res}, we perform a series of experiments to compare our various proposals.
We show that our proposed Seg-VM-Net significantly outperforms the baseline VM-Net and is more suitable for the music supervision task. 
Finally, in Section~\ref{sec:conclusion}, we discuss the limitations and perspectives of our study.
To help the reading of the paper, we summarize all the notations used in Table~\ref{tab:notation}.

\begin{figure*}
    \centering
    \includegraphics[width=\textwidth]{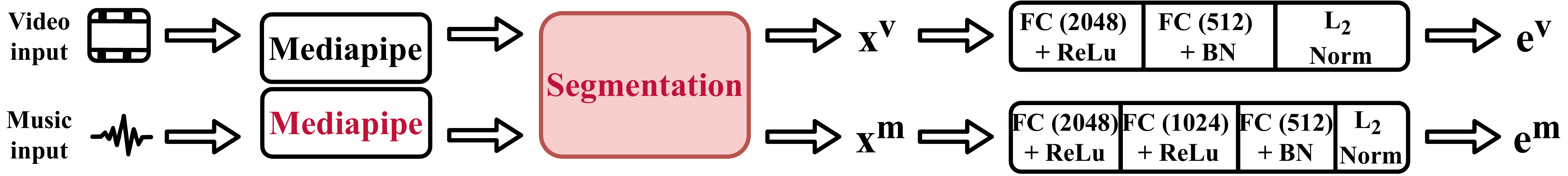}
    \caption{General overview of the studied Seg-VM-Net recommendation system (training perspective). "FC" stands for Fully-Connected and "BN" for Batch Normalization.
    We highlight in red the differences with respect to the original VM-Net, as described by Hong et al. \cite{Hong2018CBVMR:Constraint}.}
    \label{fig:train}
\end{figure*}

%%%%%%%%%%%%%%%%%%%%%%%%%%%%%%%%%%%%%%%%%%%%%%%%%%%%%%%

\section{Related work}
\label{sec:soa}

In this part, we review works related to \ac{vm} matching, i.e. the matching of video and music \textit{modalities}.
In the following, we denote by the term \textit{\ac{av} clip} an audio-video pair $c$ made of the silent video track $c^v$ and the audio music track $c^m$ extracted from the same media $c$, hence synchronized in time: $c=(c^v,c^m)$.

We also denote by the term \textit{embedding} a non-linear projection of the data in a high dimensional space. This projection aims at solving a specific downstream task (such as cross-modal recommendation).

\begin{figure}
    \centering
    \includegraphics[width=1.05\columnwidth]{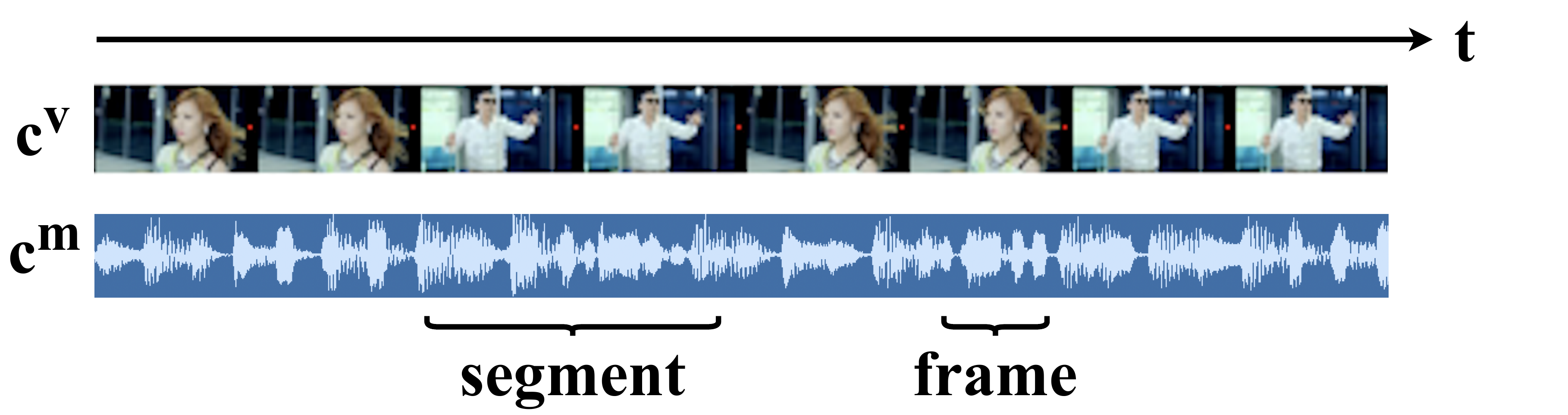}
    \caption{Illustration of the concept of \ac{av} clips and the different timescales associated: full clip, segment and frame.}
    \label{fig:related_work}
\end{figure}

% ---------------------------------
\subsection{\acl{vm} matching by semantic association}
\label{ssec:l3}

A proof-of-concept work by Liem et al.~\cite{Liem2013WhenScene} demonstrated in a user study that \ac{vm} matching could reliably be performed by users via textual descriptions.
In the machine learning literature, previous works have shown that it is also possible to automatically learn associations between audio and video files, in a supervised or self-supervised way.
The idea is to associate each element of the pair to a single representation in a multimodal space.
One representation vector can represent either the full AV clip, a segment thereof, or a short frame (typically an image and the associated 40 ms of audio).
See Figure~\ref{fig:related_work} for an illustration.

\textbf{Supervised approaches.} 
Supervised systems typically associate the audio and video modalities based on tag annotations.
Shah et al.~\cite{Shah2014ADVISORRankings} used Support Vector Machines to estimate mood tags from both modalities and then perform the final matching decision. 
Another interesting method is to project each sample directly into the valence-arousal plane via a rule-based model \cite{Sasaki2015AffectiveVideo} or a linear regression model \cite{Shin2017MusicSimilarity}.
More recently, Zeng et al.~\cite{Zeng2018Audio-VisualCCA} have developed a Deep \ac{cca} model with mood supervision for cross-modal retrieval.
While Hsia et al.~\cite{Hsia2018RepresentationRecommendation} trained a CNN to learn a semantic association of image and music via keywords extracted from the lyrics, Suris et al.~\cite{Suris2018Cross-modalRetrieval} proposed to train a fully-connected network for the audiovisual correspondence objective, regularized by a video classification task.
Similarly, Li and Kumar \cite{Li2019QueryRetrieval} trained a Convolutional Neural Network for two emotion classification tasks, one for each modality, along with the cross-modal distance learning task.
Shang et al.~\cite{Shang2021CCMR:Inputs} exploited semantic and emotion labels from lyrics to learn a connotation-aware association of music and image.
The use of external tags accelerates the training of such systems, and allows them to reach promising retrieval performances. 
However, this restricts the systems to learn a certain type of audiovisual correspondence (e.g., a mood correspondence), and requires collecting additional information to train.

\textbf{\acl{ssl} approaches.}
\ac{ssl} systems for \ac{vm} recommendation can be trained using a classification task \cite{Aytar2016SoundNet:Video}, a regression task such as source separation \cite{Zhao2018ThePixels}, or, more commonly, a metric learning task \cite{Arandjelovic2017LookLearn}.
In the latter case, a binary matching criterion is used.
The training objective is to classify pairs of audio and video pairs between "matches" (i.e. both samples are extracted from the same \ac{av} clip and from the same time position) and "non-matches". 
Early approaches have learned correlations between audio and video modalities \cite{Kuo2013BackgroundAnalysis} using multiple-type latent semantics analysis \cite{Su2013MediaEvalModeling} or dual-wing harmonium models \cite{Liao2009MiningMTV}.
Later, methods based on deep neural networks were proposed. 
With an appropriate loss function, neural networks can directly learn the correspondence between audio and video modalities. 
Aytar et al. \cite{Aytar2016SoundNet:Video} first experimented with a KL-divergence loss between the audio and video embedding vectors.
Alternatively, the L3-Net proposed by Arandjelovic and Zisserman \cite{Arandjelovic2017LookLearn} stacks a fully-connected fusion network after the two-branch embedding extraction network. 
The whole system is trained together with a binary cross-entropy loss, to predict whether both frames were extracted from the same \ac{av} clip or not.
Numerous variants of the L3-Net were proposed, for example employing curriculum learning and training with a contrastive loss \cite{Hu2020CurriculumLearning}, or using a \ac{cca} in the loss function~\cite{Yu2019DeepRetrieval}.
Adding a second stage of learning, based on distribution matching, allows to perform sounding object localization \cite{Hu2021DiscriminativeMatching}.

An interesting alternative was proposed by Owens and Efros \cite{Owens2018Audio-VisualFeatures}, whose pretext task is to predict whether audio and video excerpts extracted from the same \ac{av} clip were temporally shifted or not.
All these deep learning systems allow to leverage large non-annotated datasets of \ac{av} clips and the resulting embeddings were shown to perform well when reused for downstream tasks such as recommendation.
However, their training is computationally expensive, and the embeddings are not specific to music signals.

The main limitation of these systems is that they only associate each modality (or part thereof) to a single timeless embedding vector.
As such, they do not exploit the temporal evolution of the music and video, as we propose here.

% ---------------------------------
\subsection{Video/Music matching by sequence comparison}

Several studies suggest computing alignment metrics between each music-video pair to rank the candidates for the recommendation.

Early work by Gillet et al. \cite{Gillet2007OnVideos} performed the cross-modal recommendation task using a correlation metric between sequences of frame-level acoustic and visual features.
This work is particularly interesting in the context of music supervision, as it allows a frame-level temporal alignment between the music and video.
Here, the ranking of music tracks only relies on the variations of a low-level scalar descriptor across time.
Consequently, the semantic representation of each modality is reduced to this single descriptor.
In the context of source association, Li et al. proposed to align frame-level sequences of higher dimensional audio and visual features using the Hungarian algorithm \cite{Li2019OnlinePerformances}.
In both cases, the considered descriptors are not trained specifically for the task.
On the opposite, Wang et al. \cite{Wang2020AlignNet:Alignment} train an end-to-end, frame-level model with attention and time warping for dance-music alignment and speech-lip alignment.
However, this is only meant to align two preselected media.

To take into account the temporal evolution of each modality, another option is to model those as a sequence of segments.
For example, \cite{Lin2015EMV-matchmaker:Generation, Wang2012TheVideo, Lin2017AutomaticEditing} use correspondence of emotion sequences to compute an alignment cost between an audio and a video.
Cheng and Shen~\cite{Cheng2016OnRecommendation} represented songs and venues as sequences of concepts for context-aware music recommendation.
Gabeur et al. \cite{Gabeur2020Multi-modalRetrieval}
tackle the caption-to-video retrieval problem using a multi-modal Transformer.
Similarity at frame-level scores (a frame represents one second) are then aggregated via a learned weighted sum.

In general, the segments are defined as fixed-length excerpts and do not represent semantically meaningful units of the AV clips.
Shin et al. \cite{Shin2017MusicSimilarity} use a semantic segmentation into homogeneous units. 
The main limitation of these systems is that they are not self-supervised: they require emotion annotations.
As a consequence, they are not likely to scale to large training datasets, and they are restricted to recommending music tracks that have similar emotional content as the video.
We propose here to combine the \ac{ssl} metric learning paradigm of \cite{Aytar2016SoundNet:Video, Arandjelovic2017LookLearn, Owens2018Audio-VisualFeatures} with a segment-level approach similar to \cite{Lin2015EMV-matchmaker:Generation, Wang2012TheVideo, Lin2017AutomaticEditing, Shin2017MusicSimilarity} that takes advantage of the temporal structure of the music and video signals.

% ---------------------------------
\subsection{\acl{vm} matching by leveraging music structure}

Compared to general audio signals, music has a specific structure.
Music tracks can be divided into introductions, choruses, verses, and other semantically meaningful units, or segments.
Several studies \cite{Wang2007GenerationCues, Wang2006FullyComposition, HUA2004AutomaticAnalysis} use music structure estimation, paired with alignment techniques, to perform \textit{music video generation}.
This means that both the video and the music are preselected, and the system only performs the editing and synchronization of one of the modalities with the other.
However, there is little work about exploiting this type of variable-length segmentation for recommendation.

Our work takes inspiration from the music video generation techniques of structure-aware music segmentation and alignment of sequences.

%%%%%%%%%%%%%%%%%%%%%%%%%%%%%%%%%%%%%%%%%%%%%%%%%%%%%%%

\section{Baseline \acl{vm} recommendation system}
\label{sec:baseline}

In this part, we describe the \acf{vm} recommendation system that we will consider as our baseline.
This system, named the VM-Net, has been proposed by Hong et al. \cite{Hong2018CBVMR:Constraint}.
The VM-Net is a two-branch network where music and video are independently projected into the same embedding space, such that associated music tracks and videos are projected close to each other.
It therefore allows cross-modal recommendation: Music-to-Video or Video-to-Music.

\begin{table}
    \caption{Notations used in this paper.}
    \label{tab:notation}
    \begin{tabular}{|c|c|}
        \hline
        Variable & Definition \\
        \hline
        \hline
        $c \in \{1, \ldots, C \}$ & a specific \ac{av} clip \\
        $C$ & number of clips in the catalog \\
        $c^m$/ $c^v$ & full music/ video track of $c$ \\
        $x^m_c$/ $x^v_c$ & input music/ video feature for $c$ \\
        $e^m_c$/ $e^v_c$ & music/ video embedding for $c$ 
        \\
        $f_m(\cdot)$/ $f_v(\cdot)$ & embedding function for music/ video \\
        $f^S_m(\cdot)$/ $f^S_v(\cdot)$ & same using segmentation algorithm $S$ \\
        %\hline
        $\{c_1, \ldots, c_{K_c} \}$ & sequence of temporal segments of $c$ \\
        $K_c$ & number of segments for $c$ \\
        $\{c^{m/v}_1, \ldots, c^{m/v}_{K_c} \}$ & sequence of music/ video segments of $c$ \\
        $\{x^{m/v}_{c,1}, \ldots, x^{m/v}_{c, K_c} \}$ & input audio/ video features of the segments of $c$ \\
        $\{e^{m/v}_{c,1}, \ldots, e^{m/v}_{c, K_c} \}$ &  audio/ video embeddings of the segments of $c$ \\
        \hline
    \end{tabular}
\end{table}

% ---------------------------------
\subsection{VM-Net inputs}
\label{ssec:vmnet_inputs}

One specificity of the VM-Net is that the input of each branch is timeless, i.e. the time evolution of the track (either video or music) has been summed up as a timeless vector.
Those inputs are high-level audio and image features extracted using a previous system.
\begin{description}
\item[Audio music input: $x^m$] is a timeless feature vector of dimension 1,140.
It is obtained by statistically aggregating (using mean, max and variance values of each dimension over time) handcrafted audio features (spectral centroid, chromas, etc). The features are computed using the Librosa library\footnote{\href{https://librosa.org/doc/}{https://librosa.org/doc/}}.
\item[Video input: $x^v$] is a timeless feature vector of dimension 1,024.
To construct this vector, ImageNet features~\cite{Abu-El-Haija2016Youtube-8m:Benchmark} are computed for images of the videos sampled every second. 
The ImageNet features are then statistically aggregated using the mean values of each dimension over time.
The video ImageNet features are computed using the Mediapipe library\footnote{\href{https://github.com/google/mediapipe/tree/master/mediapipe/examples/desktop/youtube8m}{https://github.com/google/mediapipe}}.
\end{description}

% ---------------------------------
\subsection{VM-Net architecture}
\label{ssec:vmnet_archi}

The architecture of the VM-Net  (see Figure~\ref{fig:train}) is made of two independent branches: one for the music $e^m = f_{m}(x^m)$ and one for the video $e^v = f_{v}(x^v)$.
The music branch $f_m$ consists of 3 fully-connected layers of 2048, 1024, and 512 units respectively. 
The video branch $f_v$ consists of 2 fully-connected layers of 2048 and 512 units respectively.
ReLU activations are applied after each layer, except the last one. 
Linear activations and batch normalization are applied after the last layer of each branch. 
A $L_2$ normalization layer is finally applied, such that all embedding vectors have a unit norm to ensure a stable training with the triplet loss.
This architecture accounts for 6,037,504 trainable parameters and 12,084,249 floating-points operations.
The outputs of the two branches, $e^m$ and $e^v$, are considered as the music and video embeddings. 
Both are vectors of dimension 512.
Preliminary experiments validated the original architecture.

% ---------------------------------
\subsection{VM-Net training at the clip-level}
\label{ssec:vmnet_ml}

\begin{figure}
    \centering
    \includegraphics[width=0.8\columnwidth]{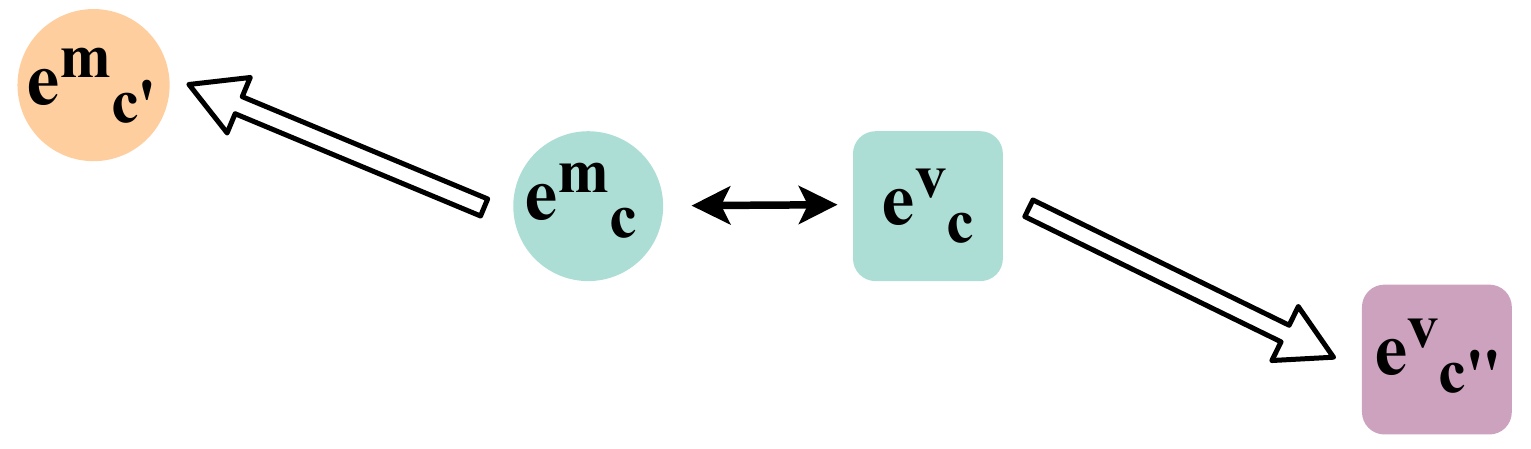}
    \caption{Illustration of the metric learning paradigm of the VM-Net. The music and video embeddings $e^m_c, e^v_c$ from a same AV clip $c$ are pulled together, while embeddings from different AV clips $c',c''$ are pushed apart.}
    \label{fig:triplet_loss}
\end{figure}

The parameters of the VM-Net are trained using metric learning in a self-supervised way.
The goal is to project the music and the video into the same multimodal embedding space, such that the music embedding $e^m$ and video embedding $e^v$ from the same \ac{av} clip are closer together than from any other sample (see Figure~\ref{fig:triplet_loss}). 
Closeness and distance are here defined in terms of Euclidean distance.

\textbf{Metric learning by triplet loss.}
To obtain this property, we train the VM-Net with an extended triplet loss named ``inter-modal ranking constraint'' as in \cite{Hong2018CBVMR:Constraint}.
The goal of a general triplet loss \cite{Weinberger2009DistanceClassification, Schroff2015FaceNet:Clustering} is to organize the projection $f$ of three samples in the embedding space such that the distance between an anchor $a$ and a positive sample $p$ is minimized, while the distance between the anchor $a$ and a negative sample $n$ is maximized.
The loss is defined as: $\mathcal{L}(a,p,n)= \max(||f(a)-f(p)||^2_2 - ||f(a)-f(n)||^2_2 + \alpha, 0)$, where $\alpha \in \mathds{R}_+^*$ is a margin parameter, and $f$ denotes the neural network embedding function.
Since \cite{Weinberger2009DistanceClassification, Schroff2015FaceNet:Clustering} deal with only one modality, there is a single $f$ function to be optimized to minimize $\mathcal{L}$.
In the case of the VM-Net, different functions $f_m$ and $f_e$ are used (for the music and video modalities) and are jointly optimized to minimize $\mathcal{L}$.
We validated the choice of the triplet loss against a classification loss in our preliminary study on the VM-Net \cite{Pretet2021DesignRecommendation}.

\textbf{\acl{ssl} at the clip level}.
In the VM-Net, the triplets are defined as such: $a$ and $p$ represent the same \ac{av} clip , while $n$ represents a different \ac{av} clip.
Let us denote by $x^v_c$ the value of $x^v$ for the clip $c \in \{1, \ldots, C\}$.
The triplet loss is then computed in both directions, using either the music or the video as anchor: $\mathcal{L}_{VM}$ with $(a,p,n) = (x^v_c, x^m_c, x^m_{c' \neq c})$ and $\mathcal{L}_{MV}$ with $(a,p,n) = (x^m_c, x^v_c, x^v_{c' \neq c})$.
To obtain the extended triplet loss, both constraints are combined via a weighted sum: $\mathcal{L}_{VMNET}= \lambda_1 \mathcal{L}_{VM} + \lambda_2 \mathcal{L}_{MV}.$
After preliminary experiments, we choose the hyperparameters $\lambda_1 = \lambda_2 = 1$ and $\alpha=0.1$.
Compared to \cite{Hong2018CBVMR:Constraint}, we choose not to use intra-modal structure constraints, as preliminary experiments showed they were not beneficial in our setup.

% ---------------------------------
\subsection{VM-Net inference at the clip-level}

To perform \acl{vm} recommendation using a clip-level system, we project each clip-level feature vector (music and video) through the corresponding branch $f_m$ and $f_v$ to get their embedding vectors.
Each \ac{av} clip $c$ is thus represented as a pair of embedding vectors: $e_c= (e^m_{c}, e^v_{c})$.

To find the most relevant music tracks for a given video query, we rank the music tracks based on the Euclidean distance between the two embedding vectors: one representing the video query $e^v_{q}$, the other representing the candidate music track of the catalog $e^m_{c}$ with $c \in \{1, \ldots, C\}$ where $C$ is the size of the catalog.

%%%%%%%%%%%%%%%%%%%%%%%%%%%%%%%%%%%%%%%%%%%%%%%%%%%%%%%

\section{Proposed \acl{v2m} recommendation system}
\label{sec:method}

One of the drawbacks of the original VM-Net system is the lack of considerations of the time evolution of the music and video modalities.
Both are summed up as timeless vectors $e_c = (e^m_c,e^v_c)$.

In this work, we introduce the Seg-VM-Net, an extension of the VM-Net which takes into account this temporal evolution.
While it is possible to consider the evolution at the frame level (either video frames or audio frames), in this work, for computational reasons,  we consider the evolution at the \textit{segment} level.
We therefore consider a \ac{av} clip $c$ as a temporal succession of non-overlapping segments $c=\{c_1, \ldots, c_{K_c}\}$, each segment being a pair of a music and a video segment: $c=\{ (c^m_{1},c^v_{1}) \ldots, (c^m_{K_c},c^v_{K_c}) \}$.
Each segment $c_k$ represents a continuous temporal region where the content is considered homogeneous either from the video content viewpoint or from the audio content viewpoint.
For example, segments can be defined as the sequence of shots in the video part $c^v$ or the sequence of verse/chorus/bridge parts in the music part $c^m$.
This naturally leads to several possible definitions of such a segmentation which we detail in the following.

% ---------------------------------
\subsection{Seg-VM-Net input music audio features $x^m$}
\label{ssec:method_features}

The original VM-Net uses handcrafted audio features, computed using the  Librosa library~\cite{McFee2015Librosa:Python}.
In preliminary experiments \cite{Pretet2021DesignRecommendation}, we showed that this choice was sub-optimal considering recent advances in feature learning.
Therefore, we replace the original handcrafted audio features of the VM-Net by pre-trained music embeddings.
We compute the music features using the Mediapipe library \cite{Gemmeke2017AudioEvents}.
As a reminder, the same library is used to extract the video ImageNet features.
Our deep feature extractor model is a VGG-inspired autotagger pre-trained on a preliminary version of the YouTube-8M dataset.
We compute the frame-level features at a rate of 1Hz and average those over time, resulting in one single 128-dimensional feature vector for each segment.
We do not re-train the feature extractors with the Seg-VM-Net, for more efficient use of computation resources.

% ---------------------------------
\subsection{Segmenting \ac{av} clips}
\label{ssec:method_segmenters}

As mentioned above, the original VM-Net considers an \ac{av} clip $c$ as a pair of timeless vectors $(x^m_c, x^v_c)$.
This accounts for a lot of information loss, as music video clips last for 3 min 54s on average in our training dataset.
In our work, we consider $c$ as a temporal succession of non-overlapping temporal segments $c=\{c_1, \ldots, c_{K_c}\}$, each segment being a pair of a music and a video segment: $c=\{ (c^m_{1},c^v_{1}) \ldots, (c^m_{K_c},c^v_{K_c}) \}$.

We define the \textit{segments} as excerpts of \ac{av} clips that are homogeneous according to one of the two modalities.
Such segmentation can therefore be obtained in various ways: for example using the sequence of shots in the video part $c^v$ or the sequence of verse/chorus/bridge parts in the music part $c^m$.
Segments can have variable duration.

We denote by $S$ a specific method to obtain such a segmentation.
Once $S$ is applied to one modality, we make the hypothesis that we can use the same segment boundaries for the other modality without severe loss of homogeneity\footnote{
This is because in the training set described in Section~\ref{ssec:vmnet_dataset}, the videos are often edited following the music tracks.}.
Since our training and test datasets are not annotated in music or video structure, $S$ is defined by a segmentation algorithm either based on the music or the video content.
We experiment with four different segmentation algorithms $S$: three music structure estimation methods and one video shot estimation method.

\begin{description}
\item[$S_{Foote}$:]
We estimate the music segment boundaries using the classic novelty-based approach introduced by \textit{Foote} \cite{Foote2000AutomaticNovelty}.
This method detects points of significant change by applying a checkerboard kernel along the diagonal of a self-similarity matrix of audio features. 
\item[$S_{SF}$:]
We estimate the music segment boundaries using the \textit{Structural Features (SF)}  algorithm introduced by Serra et al. \cite{Serra2012UnsupervisedFeatures}. 
The time series structure features are computed using a circular time-lag matrix of homogeneity and recurrence features.
Then, similar to Goto's approach \cite{Goto2006AStation.}, the novelty curve is obtained by computing the Euclidean distance between two successive columns of the lag matrix.
\item[$S_{OLDA}$:]
We estimate the music segment boundaries using the \textit{Ordinal Linear Discriminant Analysis (OLDA)} method introduced by McFee and Ellis~\cite{McFee2014LearningAnalysis}.
This supervised method adapts the linear discriminant analysis projection by only attempting to separate adjacent segments.
Then, the obtained features are clustered by merging similar successive segments.
\item[$S_{TransNet}$:]
We estimate the shot transitions using the \textit{TransNet} deep network introduced by Sou{\v{c}}ek et al~\cite{Soucek2019TransNet:Transitions}.
This convolutional neural network employs dilated 3D convolutions and is trained in a supervised way on a shot boundary detection task.
\end{description}

\textbf{Implementations.}
For the three music structure estimation algorithms (Foote, SF and OLDA), we use the implementation available in the open-source music segmentation framework \href{https://github.com/urinieto/msaf}{MSAF}~\cite{Nieto2016SystematicResearch.}.
The resulting music segments last on average 20 seconds.
For the visual shot segmentation, the open-source Python library of the TransNet shot detector was installed according to the instructions\footnote{\href{https://github.com/soCzech/TransNet}{https://github.com/soCzech/TransNet}}.
The average duration of the shots in our dataset is 6.6 seconds.

For each segmentation algorithm $S$, we segment the \ac{av} clip by considering only one modality and then apply the obtained segment boundaries to the other modality.

Since the segmentations differ according to the choice of $S$, we train the Seg-VM-Net for each choice of $S$ (using the corresponding segmented version of the training set we will describe in Section~\ref{ssec:vmnet_dataset}).
In Experiment 1 (Section~\ref{ssec:res_aggregation}) and Experiment 2 (Section~\ref{ssec:res_alignment}), we compare the performances obtained for recommendation using the different segmentation approaches $S$.

% ---------------------------------
\subsection{Seg-VM-Net training at the segment-level}
\label{ssec:method_segment_level_training}

For a given choice of $S$, each \ac{av} clip $c$ is represented as a sequence of \ac{av} segments: $c \stackrel{S}{\rightarrow} \{ c_1, \ldots c_{K_c} \}$.
For each segment $c_k$ we compute the segment-level input feature vectors $x^m_{c_k}$ and $x^v_{c_k}$ by aggregating the frame-level music and video features over the duration of the segment $c_k$ using their mean values.

For each AV clip $c$, we obtain a pair of segment-level feature vectors: $x_c=\{ (x^m_{c,1},x^v_{c,1}) \ldots, (x^m_{c,K_c},x^v_{c,K_c}) \}$. 
We then train the Seg-VM-Net using the $\mathcal{L}_{VMNET}$ triplet loss (see Section~\ref{sec:baseline}) but matching segments rather than matching clips.
This is achieved using the following triplet mining strategy that we explain for $\mathcal{L}_{VM}$.
For a given video anchor $a$ chosen as $x^v_{c,k}$ (where $c$ denotes the clip and $k$ the segment number):
\begin{itemize}
    \item the positive sample $p$ can only be the corresponding music segment from the same \ac{av} clip (same timestamp as the anchor): $x^m_{c,k}$
    \item the negative samples $n$ can only be sampled from other music tracks at any timestamp: $x^m_{c' \neq c, *}$ 
\end{itemize}
The same mining strategy is used for $\mathcal{L}_{MV}$, with $(a,p,n) = (x^m_{c,k}, x^v_{c,k}, x^v_{c'\neq c, *})$.
Note that we experimented with other mining strategies (such that allowing negatives $n$ to be sampled from the same \ac{av} clip but at different time-stamps: $n = x^m_{c,k' \neq k}$) but none had a positive impact on performance.

In practice, in each batch of size $b$ ($b=1,000$ in our case), we process $b$ different $a,p$ pairs extracted from $b$ \ac{av} clips.
For each $a,p$ pair, we use all the other samples as negatives to form $b-1$ triplets.
We then average the value of $\mathcal{L}_{VMNET}$ over all $b\times (b-1)$ available triplets.

Given that each choice of $S$ leads to a different segmentation, we train a different projection function $f_{m/v}^{S}$ for each $S$.

% ---------------------------------
\subsection{Seg-VM-Net inference at the segment-level}
\label{ssec:method_matching}

To perform \acl{vm} recommendation using our segment-level system, we first segment the query $q$ using the same segmentation algorithm $S$ as for training.
We then project each segment through the corresponding $f_{m/v}^S$ (music and video) to get the embedding vectors: $e^m_{q,k}$ and $e^v_{q,k}$ with $k \in \{1, \ldots K_q \}$.
Similarly, each \ac{av} clip $c$ of the collection is represented as a sequence of pairs of embedding vectors: $e_c=\{ (e^m_{c,1},e^v_{c,1}) \ldots, (e^m_{c,K_c},e^v_{c,K_c}) \}$.

To find the most relevant music tracks for a given video query, we rank the music tracks based on a distance $\delta$ between the two sets of embedding vectors: one representing the video query $e^v_{q,*}$, the other representing the candidate music track of the catalog $e^m_{c,*}$ with $c \in \{1, \ldots, C\}$ where $C$ is the size of the catalog.
To do so, we propose and test two families of methods: one based on cluster aggregation strategies (as used in hierarchical clustering), and one based on alignment costs.

\begin{figure}
    \centering
    \includegraphics[width=0.7\columnwidth]{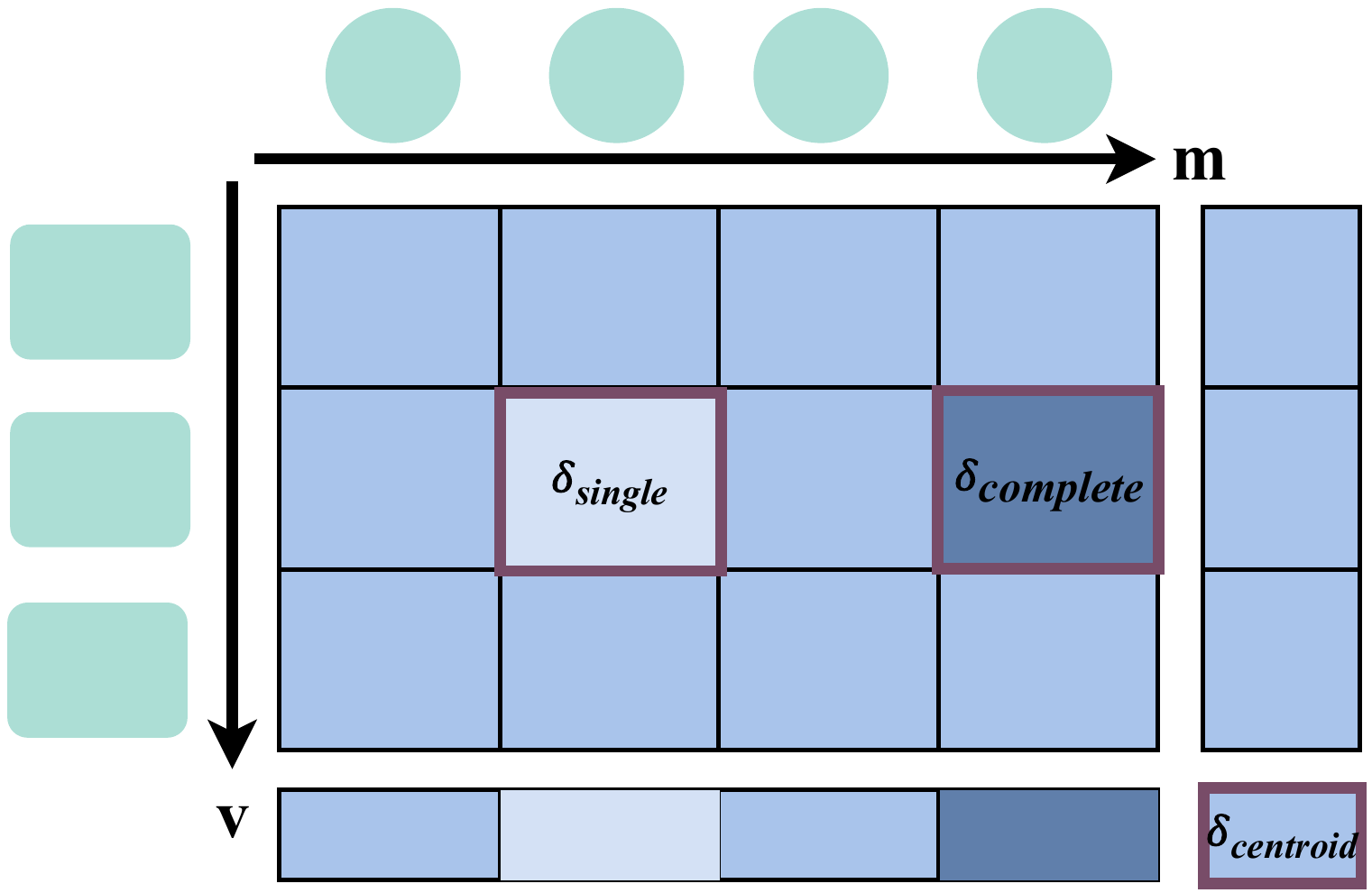}
    \caption{Experiment 1: Illustration of the proposed segment aggregation methods $\delta_{single}$, $\delta_{complete}$ and $\delta_{centroid}$.
    The darker the color of the cell, the larger the distance between the two segment embeddings. 
    }
    \label{fig:aggregation}
\end{figure}

% -----------------------------------------------------
% -----------------------------------------------------% -----------------------------------------------------
\paragraph{Cluster aggregation algorithms} 
We propose and test three simple ranking distances $\delta$ considering two unordered sets of segments.
We illustrate these different methods in Figure~\ref{fig:aggregation}, and discuss their respective performance in Experiment 1 (Section~\ref{ssec:res_aggregation}).

\begin{description}
    \item[Centroid aggregation:] 
    The distance $\delta_{centroid}$ between a music track $c$ and a video query $q$ is the Euclidean distance between the centroid of the embedding vectors of all of their respective segments. \\
    $\delta_{centroid}(c,q) = ||\frac{1}{K_c} \sum_{i=1}^{K_c} e^m_{c,i} - \frac{1}{K_q} \sum_{j=1}^{K_q} e^v_{q,j}||^2$.
    \item[Single linkage:] 
    The distance $\delta_{single}$ is the single linkage clustering between the two sets of embedding vectors. \\
    $\delta_{single}(c,q) = \min_{i,j} ||e^m_{c,i} - e^v_{q,j}||^2$.
    \item[Complete linkage:] 
    The distance $\delta_{complete}$ is the complete linkage clustering between the two sets of embedding vectors. \\
    $\delta_{complete}(c,q) = \max_{i,j} ||e^m_{c,i} - e^v_{q,j}||^2$.
\end{description}

\begin{figure*}
    \centering
    \includegraphics[width=\textwidth]{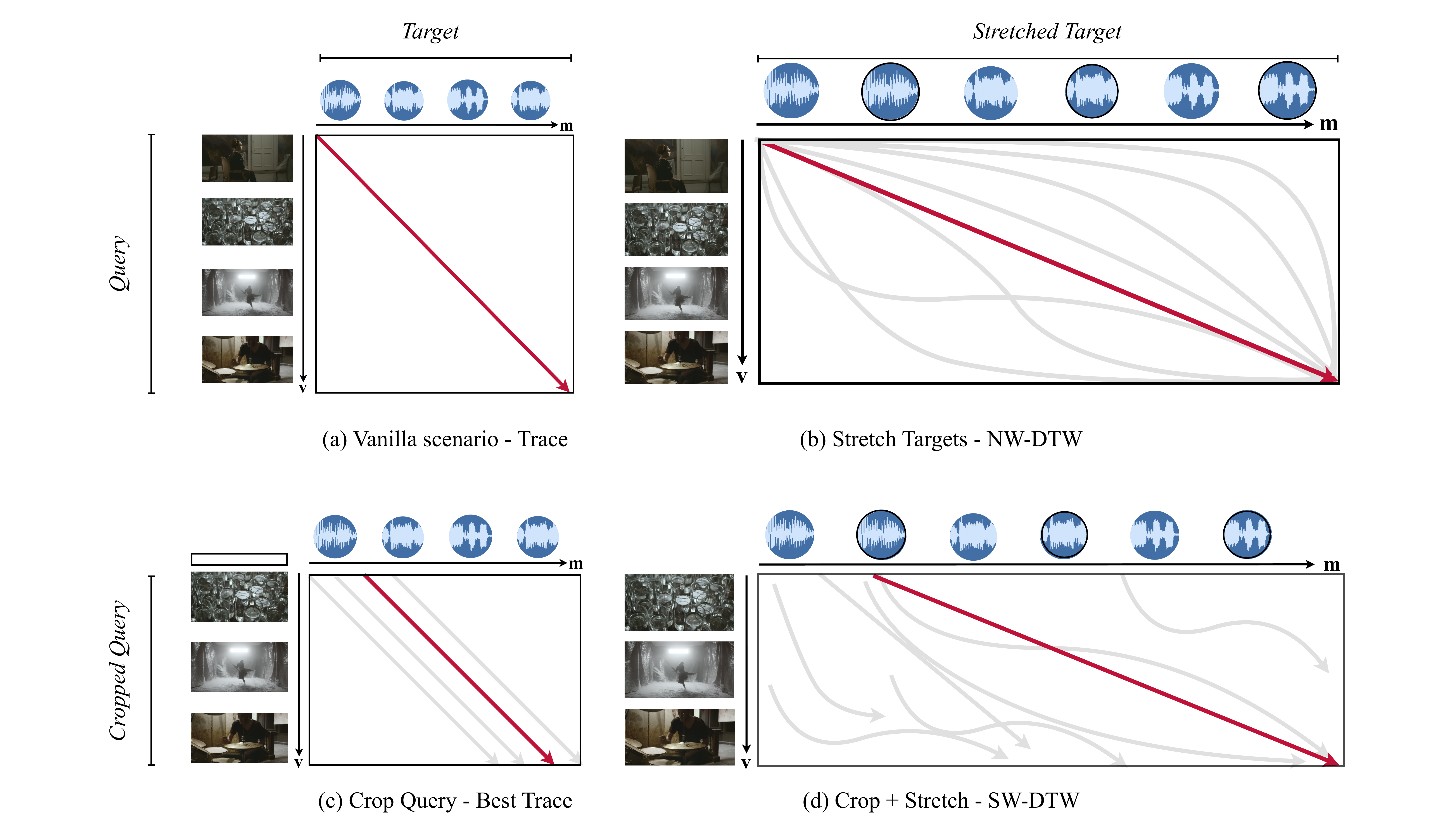}
    \caption{Schematic representations of the pairwise distance matrix between the segment embeddings: $D_{i,j} = ||e^m_{c,i} - e^v_{q,j}||^2$.
    Each subfigure represents an evaluation scenario and an example ranking distance suitable to find the optimal path (such that $\delta$ is minimal) in $D$. 
    For each evaluation scenario, we display in red an example of ground truth path. 
    We display in grey examples of paths which can be found by the represented alignment distance.}
    \label{fig:main}
\end{figure*}

\paragraph{Alignment algorithms}
A more sophisticated method is to consider the temporal succession of segments within each \ac{av} clip.
By doing so, we leverage the natural temporal ordering of the segments within each AV clip.
The ranking distances $\delta$ can then be expressed as alignment scores between two temporally ordered sequences of segments.
In Experiment 2, we investigate several ways of computing this alignment cost.
We illustrate the concept of each alignment cost on Figure~\ref{fig:main}.
We discuss their respective performance in Section~\ref{ssec:res_alignment}.

\begin{description}
    \item[NW-DTW:] 
    The distance $\delta_{NW-DTW}$ is a \ac{DTW} cost between the two sequences of embedding vectors. 
    The DTW algorithm used is inspired from the Needleman-Wunsch algorithm~\cite{Needleman1970AProteins}. 
    We use an \textit{indel} cost of $0.05$, determined via a grid search. 
    The NW-DTW we use is detailed in Algorithm~\ref{nw-dtw} for the case of similarity (the distance is then the opposite) where $m[i]=e^m_{c,i}$ and $v[j]= e^v_{q,j}$.
    \item[SW-DTW:] 
    The distance $\delta_{SW-DTW}$ is a \ac{DTW} cost between the two optimal subsequences of embedding vectors. 
    The DTW algorithm used is inspired by the Smith-Waterman algorithm~\cite{Smith1981IdentificationSubsequences}. 
    We use an \textit{indel} cost of $0.01$, determined via a grid search. 
    Contrary to NW-DTW, the SW-DTW variant allows to compute the alignment cost based on local sequence alignment rather than on a full path.
    The SW-DTW we use is detailed in Algorithm~\ref{sw-dtw} for the case of similarity (the distance is then the opposite) where $m[i]=e^m_{c,i}$ and $v[j]= e^v_{q,j}$.
    \item[Trace:] 
    $\delta_{trace}$ is the sum of the coefficients on the main diagonal of the pairwise distance matrix:
    \\ $\delta_{trace}(c,q) = \sum_{i=1}^{min(K_c,K_q)} ||e^m_{c,i} - e^v_{q,i}||^2$.
    \item[Best (B.) Trace:] 
    $\delta_{btrace}$ is the minimal trace over all diagonals of the pairwise distance matrix:
\end{description}

\begin{center}
$\delta_{bt}(c,q) = \min\limits_{k \in [0,K_q-K_c]} \sum_{i=1}^{K_c} ||e^m_{c,i} - e^v_{q,i+k}||^2$ if $K_c<K_q$;
\end{center}
\begin{center}
$\delta_{bt}(c,q) = \min\limits_{k \in [0,K_c-K_q]} \sum_{i=1}^{K_q} ||e^m_{c,i+k} - e^v_{q,i}||^2$ if $K_c>K_q$.
\end{center}
    
\begin{algorithm}[t]
    \SetAlgoLined
    \KwResult{NW-DTW($m,v$)}
    $indel \gets 0.05$\;
    \For{$i\gets0$ \KwTo $len(m)$}{
    $X[i, 0]\gets -indel*i$\;
    }
    \For{$j\gets0$ \KwTo $len(v)$}{
    $X[0, j]\gets -indel*j$\;
    }
    \For{$i\gets 1$ \KwTo $len(m)$}{
        \For{$j\gets 1$ \KwTo $len(v)$}{
        $sim\gets m[i] \cdot v[j]$\;
        $X[i, j]\gets \max(X[i-1, j]-indel, X[i, j-1]-indel, X[i-1, j-1]+sim)$\;
    }
    }
    \Return $X[len(m), len(v)]$\;
 \caption{Needleman-Wunsch DTW}
 \label{nw-dtw}
\end{algorithm}

\begin{algorithm}[t]
    \SetAlgoLined
    \KwResult{SW-DTW($m,v$)}
    $indel \gets 0.01$\;
    \For{$i\gets0$ \KwTo $len(m)$}{
    $X[i, 0]\gets 0$\;
    }
    \For{$j\gets0$ \KwTo $len(v)$}{
    $X[0, j]\gets 0$\;
    }
    \For{$i\gets 1$ \KwTo $len(m)$}{
        \For{$j\gets 1$ \KwTo $len(v)$}{
        $sim\gets m[i] \cdot v[j]$\;
        $X[i, j]\gets \max(0, X[i-1, j]-indel, X[i, j-1]-indel, X[i-1, j-1]+sim)$\;
    }
    }
    \Return $\max(X)$\;
 \caption{Smith-Waterman DTW}
 \label{sw-dtw}
\end{algorithm}

%%%%%%%%%%%%%%%%%%%%%%%%%%%%%%%%%%%%%%%%%%%%%%%%%%%%%%%

\section{Evaluation protocol}
\label{sec:eval}

In this part, we describe the evaluation protocol we used for our experiments which aim at comparing the performances of our various proposals to the ones of the original VM-Net.

% ---------------------------------
\subsection{Training details}
\label{ssec:vmnet_training}

We reimplemented the VM-Net in Keras from the provided Tensorflow code\footnote{\href{https://github.com/csehong/VM-Net}{https://github.com/csehong/VM-Net}}.
We then use the ADAM optimizer with a learning rate of $10^{-6}$, and a batch size of 1000. 
A dropout scheme with a probability of 0.5 and early stopping are applied to prevent overfitting.
All experiments are performed on an Nvidia GeForce GTX 1080 Ti GPU.

% ---------------------------------
\subsection{Training set: HIMV-50K}
\label{ssec:vmnet_dataset}

To train the VM-Net, we use a subset of the YouTube-8M dataset~\cite{Abu-El-Haija2016Youtube-8m:Benchmark}.
This subset corresponds to 51,000 \ac{av} clips annotated as ``music video''.
We partition this subset in 50,000 randomly selected pairs for training,
and the remaining 1,000 for validation.
In the following, we denote this dataset as HIMV-50K.
Although the HIMV-50K dataset contains no labels, its large size makes it suitable for self-supervised training.
It should be noted that while the audio of these \ac{av} clips always contains music, the video can be anything from professional promotional videos to amateur montages of still images.

% ---------------------------------
\subsection{Test set: MVD}
\label{ssec:eval_dataset}

While the HIMV-50K dataset is large and diverse (and therefore suitable for training a neural network), the quality of the video part of the \ac{av} clips is very heterogeneous.
Evaluating the Seg-VM-Net on this noisy dataset would make results difficult to interpret.
For this reason, we use a cleaner dataset to evaluate the performance of our system: the Music Video Dataset (MVD)~\cite{Schindler2016HarnessingRetrieval, Schindler2019Multi-ModalAnalysis}.
The MVD consists in 2,212 music video clips, manually curated. 
The music and video parts of the clips are of professional quality.
The average duration of each \ac{av} clip is 4 minutes.
We randomly select $N=1,000$ of these clips to evaluate our systems.

% ---------------------------------
\subsection{Vanilla evaluation scenario}
\label{ssec:eval_metrics}

The real use case scenario would consist in finding the most appropriate music track (here denoted as target) $t^{m}$ from a catalog given a video query $q^v$; this independently of the respective duration of $t^{m}$ and $q^v$.
However, the evaluation of this real use case scenario would imply human evaluation which cannot be done at scale.
To allow performing the evaluation at scale (i.e. using 1,000 different queries), we therefore create two evaluation scenarios.

In the vanilla evaluation scenario, we consider in turn each of the 1,000 clips.
For each clip $c$, we use its video part $c^v$ as a query: $q^v=c^v$.
The goal is to retrieve from the whole set of clips the corresponding music part, i.e. $t^{m}=c^m$.

Although convenient (as it allows a direct evaluation of our systems), this scenario is obviously not realistic, since as $q^v$ and $t^{m}$ are coming from the same \ac{av} clip $c$, they share the exact same boundaries $\{c_1, \ldots c_{K_c} \}$. 
Their segments are therefore perfectly aligned (See Figure~\ref{fig:main} (a)).
To solve this issue, we propose a second evaluation scenario, denoted by ``realistic" which mimics the real use case by applying artificial modifications (temporal cropping, stretching) to $q^v$ or $t^{m}$ (see part~\ref{ssec:eval_robustness} for details).

In both cases, to obtain a ranked list of recommendations for a given video query, we compute the pairwise distances $\{\delta(c^m,q^v)\}_{c \in \{1,\ldots,C \}}$ between the query video embedding $e^v_q$ and all music track embeddings $\{e^m_{c}\}_{c \in \{1,\ldots,C \}}$. 
This ranking distance $\delta$ depends on the experiment.
We then rank all music tracks by increasing distance $\delta$.
Please note that all distances studied in this paper are symmetrical, and could therefore be used to retrieve videos from a music query.
For simplification purposes, and in line with the music supervision use case, we choose here to focus on the \ac{v2m} case.

% ---------------------------------
\subsection{Realistic evaluation scenario}
\label{ssec:eval_robustness}

In the music supervision task, the best soundtrack is not known a priori.
It may also have a different number of segments than the video query, thus requiring some editing.
Therefore, the vanilla evaluation scenario only accounts for very specific cases.
A relevant recommendation system for music supervision should be able to find the best music track from the catalog, even if some transformations are required in order to obtain the optimal fit.
In other words, we need our recommendation system to be robust against video queries that are not perfectly aligned to the candidate music tracks. 

In Experiment 3, we test the robustness of the different ranking distances $\delta$ using more realistic evaluation scenarios.
Since we do not have realistic ground truth data, we apply three types of perturbations to the queries and targets in order to imitate real-life use cases:
\begin{description}
\item["Crop Query"]: we remove the first two segments of each video query: $\{c^v_1, c^v_2, c^v_3, ..., c^v_{K_c}\} \rightarrow \{c^v_3, ..., c^v_{K_c}\}$ (See Figure~\ref{fig:main} (c)).
\item["Stretch Targets"]: we repeat each segment of the target music tracks: $\{c^m_1, ..., c^m_{K_c}\} \rightarrow \{c^m_1, c^m_1, ..., c^m_{K_c}, c^m_{K_c}\}$ (See Figure~\ref{fig:main} (b)).
\item["Crop+Stretch"]: we apply simultaneously the "Crop Query" and "Stretch Targets" perturbations (See Figure~\ref{fig:main} (d)).
\end{description}

We use these simple transformations in line with the music supervision needs.
Indeed, usual video queries (e.g., commercials) are often shorter than full-length music tracks.
One common edit on the music tracks is therefore to crop them to select the most relevant excerpt.
Swapping segments would be much less common, according to the professionals interviewed by the authors (see \cite{Pretet2021IsStudy} for more details).

As before, we compute all pairwise alignment costs with the modified queries or targets to obtain the rankings.
We discuss the performance of our system against each evaluation scenario in Section~\ref{ssec:res_best}.

% ---------------------------------
\subsection{Performance measures}
\label{ssec:eval_measures}

In both evaluation scenarios (vanilla or realistic), there is a single correct music track to be retrieved among the 1,000 of the dataset.

If the corresponding music is in the top $k$ of the ranked list, we set the Recall at $k$ ($R@k$) to 1, otherwise to 0. 
For each query, $R@k$ is hence binary. Higher is better.
We also compute the $Rank$ at which the target music appears in the ranked list. 
If it is displayed first, we set the $Rank$ to 1, if it is the last recommendation, the $Rank$ is equal to 1,000. Lower is better.
We repeat this operation using all videos of the test set as the query $q^v$.
Since the test set MVD is different from the training set HIMV-50K, none of the test clips (neither query nor targets) were seen during training.

The two final metrics are: the average of $R@k$ over the 1,000 test clips, displayed as percentages (we use $k \in \{1,10,25\}$) and the mean $Rank$ over the 1,000 test clips.
We display the confidence interval at 95\% for the $Rank$.
We do not display confidence intervals for $R@k$, as it is a binary metric, and its distribution cannot be considered Gaussian.

%%%%%%%%%%%%%%%%%%%%%%%%%%%%%%%%%%%%%%%%%%%%%%%%%%%%%%%

\section{Evaluation results}
\label{sec:res}

In each experiment, we trained the systems on 50,000 \ac{av} pairs from the HIMV-50K dataset, with validation on the 1,000 remaining pairs from the HIMV-50K. Evaluation was performed on a different unseen dataset using 1,000 pairs from the MVD.
The training time of each of the four Seg-VM-Nets is approximately four days.

% ---------------------------------
\subsection{Experiment 1: Segment clustering aggregation algorithms}
\label{ssec:res_aggregation}

In the first experiment, we compare the performances of the original \textit{clip-level} VM-Net  (represented by the baseline systems of \cite{Hong2018CBVMR:Constraint} and \cite{Pretet2021DesignRecommendation}) to our proposed \textit{segment-level} Seg-VM-Nets.
All evaluations are performed using the vanilla scenario.

\textbf{Clip-level baseline systems.}
We consider two baseline systems trained at \textit{clip level} (see Sections~\ref{ssec:vmnet_inputs} and \ref{ssec:vmnet_ml}).
The two only differ by the choice of the input audio features.
The Baseline~1 is the VM-Net trained at clip level with the original handcrafted audio features obtained from Librosa, as in \cite{Hong2018CBVMR:Constraint}.
The Baseline~2 uses pre-trained music features obtained from Mediapipe, as we proposed in \cite{Pretet2021DesignRecommendation}.
These Mediapipe features are the ones we use in all our segment-level systems.

\textbf{Segment clustering aggregation algorithms.}
We compare the three clustering aggregation algorithms, each using all four segmentation algorithms $S$: based on the music structure ($S_{Foote}, S_{SF}, S_{OLDA}$) or based on the video structure ($S_{TransNet}$).
$S$ is used to define the segments which are used 
(a) at training time (leading to different embedding projections $e=f_{m/v}^S(x)$), and
(b) at inference time to match the segments of two clips.
For the inference, we compare the use of the three  cluster aggregation methods: $\delta_{centroid}$, $\delta_{single}$ and $\delta_{complete}$.
In this experiment, the matching does not consider the temporal organization of the segments.

\begin{table}
\caption{Results of Experiment 1 in terms of Mean $Rank$: comparing segmentation methods $S$ with clustering aggregation algorithms $\delta$ in the vanilla evaluation scenario. 
Lower is better. 
$S_{Foote}$ and $\delta_{centroid}$ perform best.}
\label{res1}
     \centering
     \begin{tabular}[width=\columnwidth]{|c|c|c|c|c|}
         \hline
         & Centroid & Single Linkage & Complete Linkage \\ 
         \hline
         Foote & \textbf{94 $\pm$ 17} & 120 $\pm$ 18 & 324 $\pm$ 37 \\
         SF & 100 $\pm$ 17 & 125 $\pm$ 18 & 337 $\pm$ 36 \\
         OLDA & \textbf{94 $\pm$ 16} & 128 $\pm$ 19 & 347 $\pm$ 36 \\
         TransNet & 113 $\pm$ 1 & 172 $\pm$ 22 & 283 $\pm$ 31 \\
        \hline
         Baseline 1 \cite{Hong2018CBVMR:Constraint} & \multicolumn{3}{c|}{193 $\pm$ 25} \\ 
         Baseline 2 \cite{Pretet2021DesignRecommendation} & \multicolumn{3}{c|}{118 $\pm$ 19 } \\ 
         Chance & \multicolumn{3}{c|}{500} \\ 
         \hline
     \end{tabular}
 \end{table}

 \begin{figure}
     \centering
     \includegraphics[width=\columnwidth]{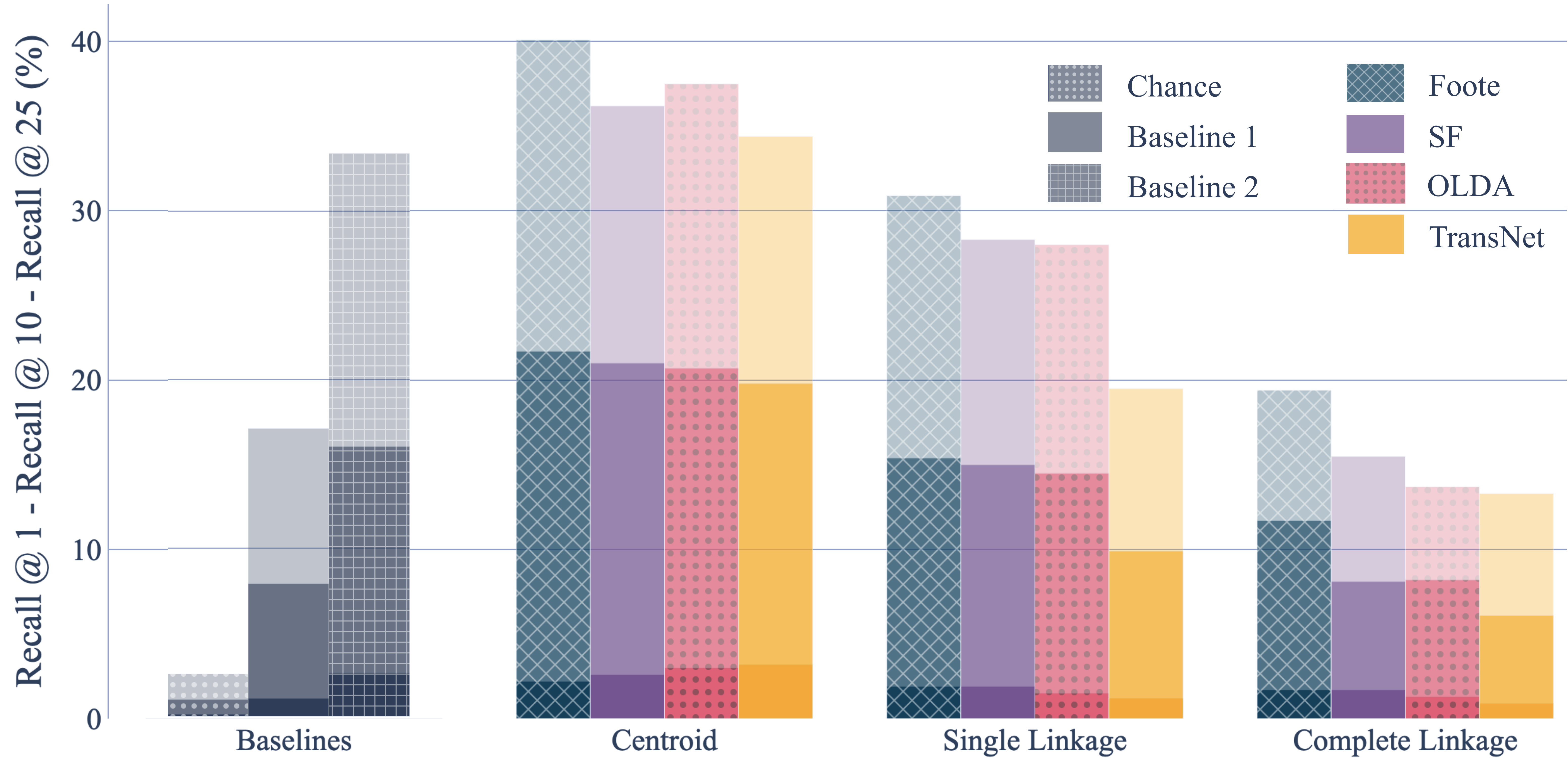}
     \caption{Results of Experiment 1 in terms of Recall: comparing segmentation methods $S$ with clustering aggregation techniques $\delta$ in the vanilla evaluation scenario. 
     The three Recall metrics ($R@1$, $R@10$ and $R@25$) are stacked with decreasing opacity. 
     Higher is better. 
     $S_{Foote}$ performs best for all alignment methods. 
     Best viewed in color.}
     \label{fig:barchart1}
 \end{figure}

Table~\ref{res1} and Figure~\ref{fig:barchart1} show the results of Experiment 1.

\textbf{Computation time.}
The training time of the Seg-VM-Net for this experiment varies between two and six days.
Inference took less than one minute to process the 1,000 test queries.

\textbf{Among the segment clustering aggregation methods, the Centroid method performs best for almost all metrics.}
$\delta_{centroid}$ perform best, while $\delta_{complete}$ performs worst.
$\delta_{centroid}$ trained on segments defined by $S_{Foote}$ performs better than the clip-level baselines on all metrics.
To understand this, let us be reminded that the clip-level systems are trained on input vectors $x^m_c$/$x^v_c$ which represent the whole \ac{av} clip. 
On the opposite end, the segment-level systems are trained on input vectors which represent segments $x^m_{c,k}$/$x^v_{c,l}$ of the \ac{av} clips of \textit{variable duration}.
In $\delta_{centroid}$, the resulting embeddings are then averaged over the clip to produce the recommendation.
As a result, for $\delta_{centroid}$, short and long segments have the same weight for the recommendation.
We expect this to be an advantage in the case where shorter segments are especially meaningful.

\textbf{We observe a variation of the performance of the Seg-VM-Net according to the segmentation strategy $S$ of the \ac{av} clips.}
The segmentation $S_{Foote}$, which is based on homogeneity of consecutive audio frames, seems to perform best for all the cluster aggregation methods.
$S_{TransNet}$ results in poor performance in almost every setting.
Compared to the three audio-based methods, $S_{TransNet}$ tends to produce many more segments per clip (i.e. the average shot duration is shorter than the average music segment duration).
While most audio boundaries are reflected by cuts when editing a music video, we do not expect every cut to have a signification in terms of music.
This may explain why $S_{TransNet}$ has poor performance.

Experiment 1 therefore supports the intuition that the segment-level embeddings carry extra representation power, but a better ranking distance $\delta$ is required in order to take full advantage of it.

% ---------------------------------
\subsection{Experiment 2: Segment alignment algorithms}
\label{ssec:res_alignment}

\begin{figure}
     \centering
     \includegraphics[width=\columnwidth]{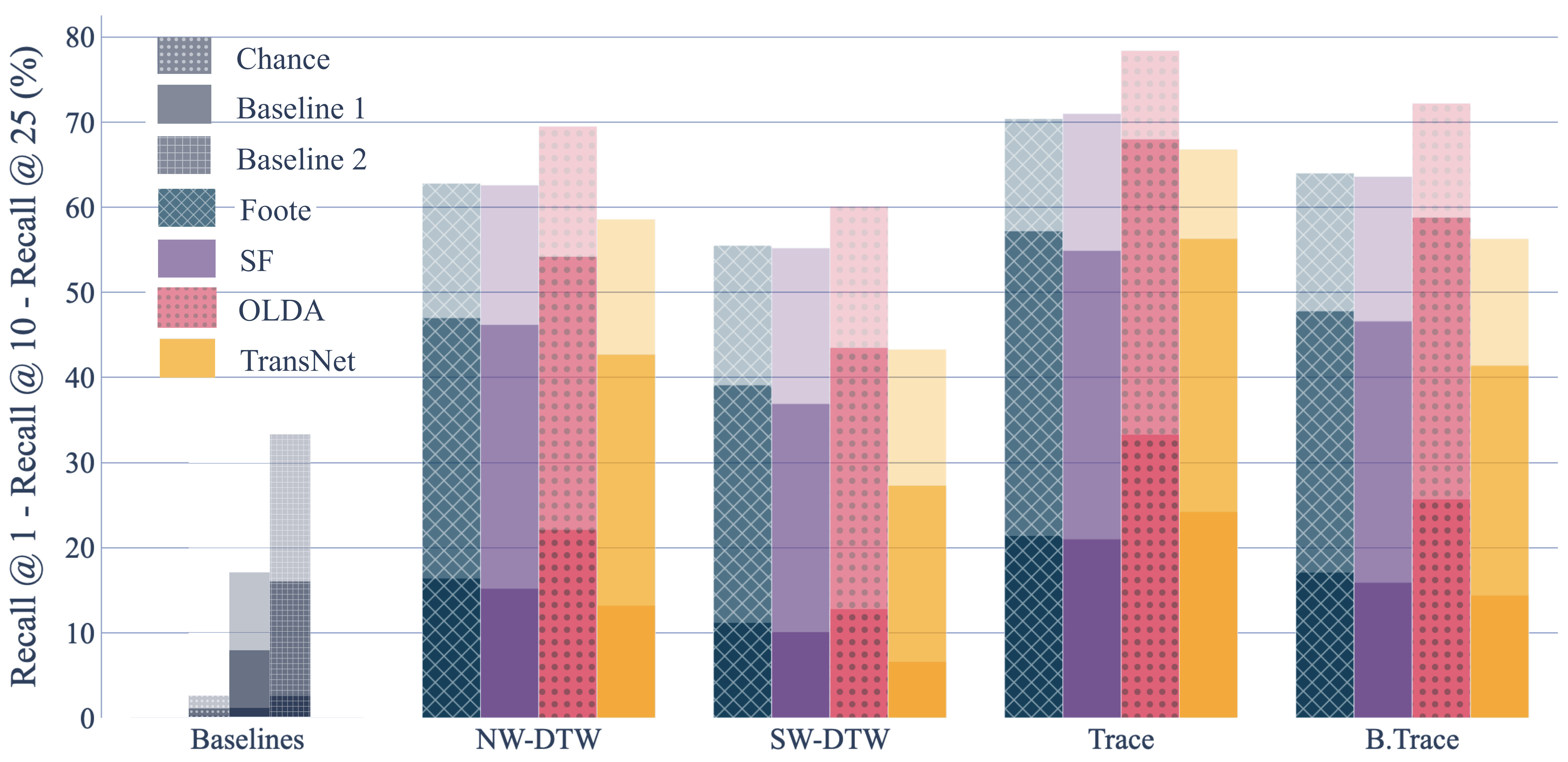}
     \caption{Results of Experiment 2 in terms of Recall: comparing segment alignment algorithms in the vanilla evaluation scenario. 
     The three Recall metrics ($R@1$, $R@10$ and $R@25$) are stacked with decreasing opacity. 
     Higher is better. 
     $S_{OLDA}$ performs best for all alignment algorithms. 
     Best viewed in color.}
     \label{fig:barchart2}
 \end{figure}

In the second experiment, we focus on the segment-level systems.
We compare the performance obtained using the four sequence alignment algorithms $\delta$: $\delta_{NW-DTW}$, $\delta_{SW-DTW}$, $\delta_{trace}$ and $\delta_{btrace}$.
We also still compare the segmentation algorithms $S_{Foote}, S_{SF}, S_{OLDA}$ and $S_{TransNet}$.
All evaluations are still performed using the vanilla scenario.
Table~\ref{res2} and Figure~\ref{fig:barchart2} show the results of Experiment 2.

\textbf{Segment alignment metrics.} 
It seems clear that the segment-level approach combined with alignment metrics allows for a large improvement compared to the clip-level systems but also compared to the cluster aggregation methods (Experiment 1).
According to Figure~\ref{fig:barchart2}, it seems that this time $S_{OLDA}$ allows for the largest improvement.
The highest $R@25$ is achieved using $\delta_{Trace}$ and $S_{OLDA}$: the Seg-VM-Net then reaches a $R@25=78.40$ (for more than three out of four queries, the Seg-VM-Net was able to retrieve the matching music track in its top 25). 
Table~\ref{res2} indicates that for this configuration, the Seg-VM-Net achieves a Mean $Rank$ of 32 (the matching video is on average at rank 32, out of 1,000 videos).
In comparison, the Baseline~2 clip-level system had a much lower $R@25$ (33.4) and much higher Mean $Rank$ (118).
The improvement is also very noticeable compared to the $\delta_{centroid}$ segment-level (see Experiment 1) which had a $R@25=40.10$.

The fact that $\delta_{trace}$ provides the best results in the vanilla evaluation scenario is not surprising, since in this scenario, the audio and video segments perfectly match each other.
For the vanilla scenario, there is no gain to use the more elaborated $\delta_{NW-DTW}$, $\delta_{SW-DTW}$ and $\delta_{btrace}$.
Performing a global alignment ($\delta_{NW-DTW}$) or allowing an offset to the trace ($\delta_{btrace}$) both result in degraded performance.
The subsequence alignment method ($\delta_{SW-DTW}$) produces the worst performance.

\begin{table}
\caption{Results of Experiment 2 in terms of Mean $Rank$: comparing segment alignment algorithms in the vanilla evaluation scenario. Lower is better. 
$\delta_{trace}$ performs best both in terms of metrics and inference duration.}
\label{res2}
     \centering
     \scriptsize
     \begin{tabular}[width=\columnwidth]{|c|l|c|c|}
         \hline
         & \multicolumn{1}{c|}{} & \multicolumn{1}{c|}{Mean Rank} & \multicolumn{1}{c|}{Time} \\
         \hline
         & Chance & 500 & 00'00'' \\
         & Baseline 1 \cite{Hong2018CBVMR:Constraint} & 193 $\pm$ 25 & 00'10'' \\
         & Baseline 2 \cite{Pretet2021DesignRecommendation} & 118 $\pm$ 19 & 00'10'' \\
         \hline
         \parbox[t]{2mm}{\multirow{4}{*}{\rotatebox[origin=c]{90}{NW-DTW}}} 
         & Foote & 56 $\pm$ 13 & 11'59''\\
         & SF & 56 $\pm$ 13 & 10'09'' \\
         & OLDA & 46 $\pm$ 12 & 13'41'' \\
         & TransNet & 60 $\pm$ 13 & 48'29'' \\
         \hline
         \parbox[t]{2mm}{\multirow{4}{*}{\rotatebox[origin=c]{90}{SW-DTW}}} 
         & Foote & 68 $\pm$ 14 & 19'40'' \\
         & SF & 70 $\pm$ 15 & 16'20'' \\
         & OLDA  & 62 $\pm$ 14 & 21'14'' \\
         & TransNet & 99 $\pm$ 17 & 1h13'56'' \\
         \hline
         \parbox[t]{2mm}{\multirow{4}{*}{\rotatebox[origin=c]{90}{Trace}}} 
         & Foote & 44 $\pm$ 12 & 00'55'' \\
         & SF & 44 $\pm$ 11 & 01'00'' \\
         & OLDA & \textbf{32 $\pm$ 10} & \textbf{00'53''} \\
         & TransNet & 43 $\pm$ 11 & 01'45'' \\
         \hline
         \parbox[t]{2mm}{\multirow{4}{*}{\rotatebox[origin=c]{90}{B. Trace}}} 
         & Foote & 55 $\pm$ 13 & 01'45'' \\
         & SF & 55 $\pm$ 13 & 01'32'' \\
         & OLDA & 43 $\pm$ 12 & 01'35'' \\
         & TransNet & 73 $\pm$ 16 & 03'51'' \\
         \hline
     \end{tabular}
 \end{table}

 \textbf{Computation time.}
In this experiment, inference time varies a lot depending on the choice of $\delta$; we provide each inference time in the table of results.
We performed all inferences on an Intel Core i7-6850K CPU with 64G of RAM.
Overall, the inference duration of $\delta_{trace}$ and $\delta_{btrace}$ seem reasonable to us, as it represents the processing of 1,000 queries.
In a realistic use case, the user will enter one query at a time, thus expecting a response in about 0.1s for the system trained on $S_{OLDA}$ segments with the $\delta_{trace}$ alignment cost.
Using the $S_{TransNet}$ segmentation results systematically in longer inference time, because of the higher number of segments to align.
Both \ac{dtw} alignment methods ($\delta_{NW-DTW}$ and $\delta_{SW-DTW}$) result in high computation times.
This was expected, as their complexity is quadratic.
Only allowing an offset ($\delta_{btrace}$) results in a slight increase of computation time compared to $\delta_{trace}$ (1'35 instead of 53'' for OLDA), with performance similar to the more costly $\delta_{NW-DTW}$. 

% ---------------------------------
\subsection{Experiment 3: Realistic evaluation}
\label{ssec:res_best}

While Experiments 1 and 2 were achieved using the vanilla evaluation scenario, we now test the realistic evaluation scenario where query and targets do not (necessarily) have the same number of segments.
To mimic the real-case scenario, we artificially modified either the query or the targets (using ``Crop Query'', ``Stretch Targets'' and ``Crop+Stretch'').

We then again compare the various pairwise ranking distances $\delta_{centroid}$, $\delta_{NW-DTW}$, $\delta_{SW-DTW}$, $\delta_{trace}$ and $\delta_{btrace}$ but using only the best segmentation algorithm $S_{OLDA}$.

\begin{table}
\caption{Results of Experiment 3 in terms of Mean $Rank$ using the realistic evaluation and $S_{OLDA}$. 
Lower is better. 
For the scenario ``Crop Query'', $\delta_{btrace}$ is the most robust.
For the scenarios ``Stretch Targets'' and ``Crop+Stretch'', $\delta_{NW-DTW}$ is the most robust.}
\label{res3}
\centering
\begin{tabular}{|c|c|c|c|}
    \hline
    & Crop Query & Stretch Targets & Crop+Stretch \\
     \hline
     Centroid & 96 $\pm$ 16 & 94 $\pm$ 16 & 96 $\pm$ 16 \\ 
     NW-DTW & 75 $\pm$ 16 & \textbf{55 $\pm$ 12} & \textbf{65 $\pm$ 13} \\ 
     SW-DTW & 72 $\pm$ 15 & 68 $\pm$ 15 & 79 $\pm$ 16 \\ 
     Trace & 92 $\pm$ 17 & 90 $\pm$ 16 & 146 $\pm$ 22 \\ 
     Best Trace & \textbf{60 $\pm$ 14} & 83 $\pm$ 15 & 89 $\pm$ 16 \\ 
     \hline
\end{tabular}
\end{table}

 \begin{figure}
     \centering
     \includegraphics[width=\columnwidth]{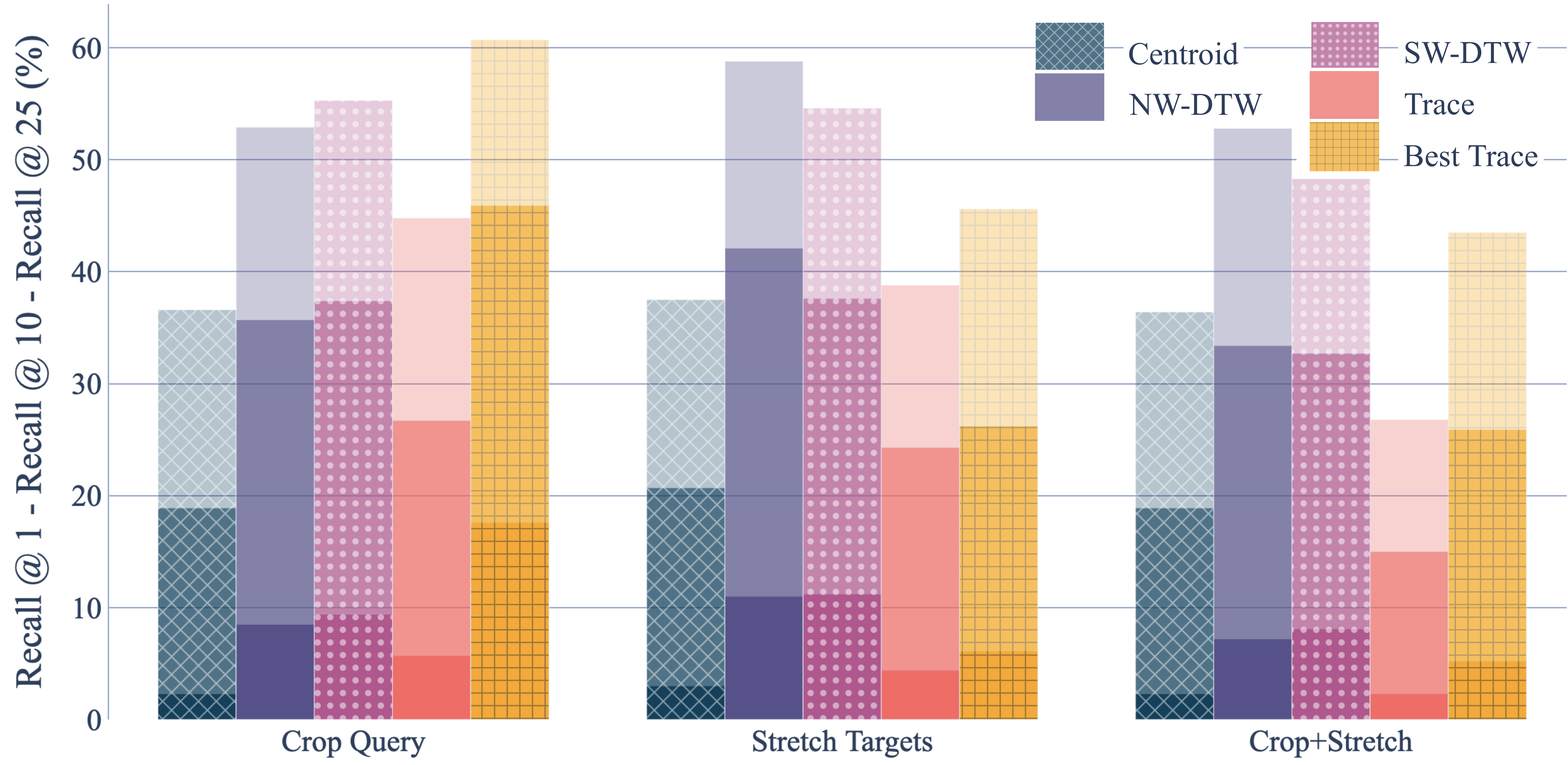}
     \caption{Results of Experiment 3 in terms of Recall using the realistic evaluation scenario and $S_{OLDA}$. 
     The three Recall metrics ($R@1$, $R@10$ and $R@25$) are stacked with decreasing opacity. 
     Higher is better. 
     For the scenario ``Crop Query'', $\delta_{btrace}$ performs best.
     For the scenarios ``Stretch Targets'' and ``Crop+Stretch'', $\delta_{NW-DTW}$ is the most robust.
     Best viewed in color.}
     \label{fig:barchart3}
 \end{figure}

Table~\ref{res3} and Figure~\ref{fig:barchart3} show the results of Experiment 3.
As can be seen, the perturbations of the queries and targets result in a degradation of the performance of all systems, compared to the vanilla evaluation procedure (see Table~\ref{res2}).
The $\delta_{centroid}$ and $\delta_{trace}$ systems are particularly impacted by this degradation.
Following our intuition, $\delta_{btrace}$ is the most robust against the ``Crop Query'' perturbation.
Additionally, $\delta_{NW-DTW}$ and $\delta_{SW-DTW}$ are the most robust against the ``Stretch Targets'' and ``Crop+Stretch'' degradations.
This was expected, as the DTW alignment algorithms are the only ones that can handle stretched alignment (See Figure~\ref{fig:main}).
The $\delta_{NW-DTW}$, which is constrained to use all segments from the first to the last one, performs best in terms of $R@10$, $R@25$ and Rank.
The $\delta_{SW-DTW}$, which can ignore segments at the beginning and end of the media, performs best in terms of $R@1$.

% ---------------------------------
\subsection{Qualitative analysis}
\label{ssec:res_demo}

\begin{figure*}
    \centering
    \includegraphics[width=\textwidth]{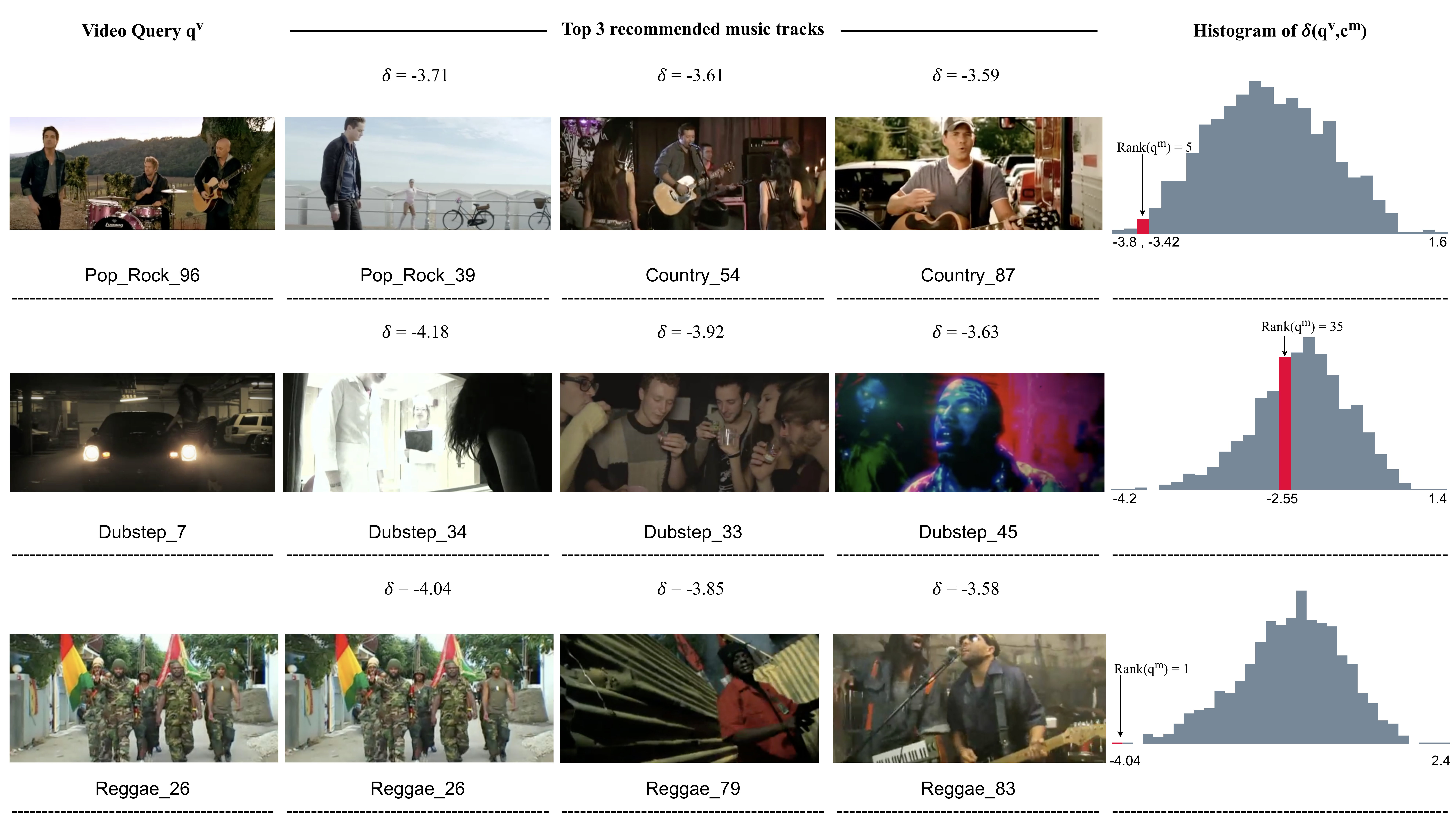}
    \caption{Qualitative analysis of the results of the system which uses $S_{OLDA}$ segmentation with $\delta_{trace}$ segment aggregation algorithm; for three video queries from the MVD.}
    \label{fig:demo}
\end{figure*}

\begin{figure}
    \centering
    \includegraphics[width=0.8\columnwidth]{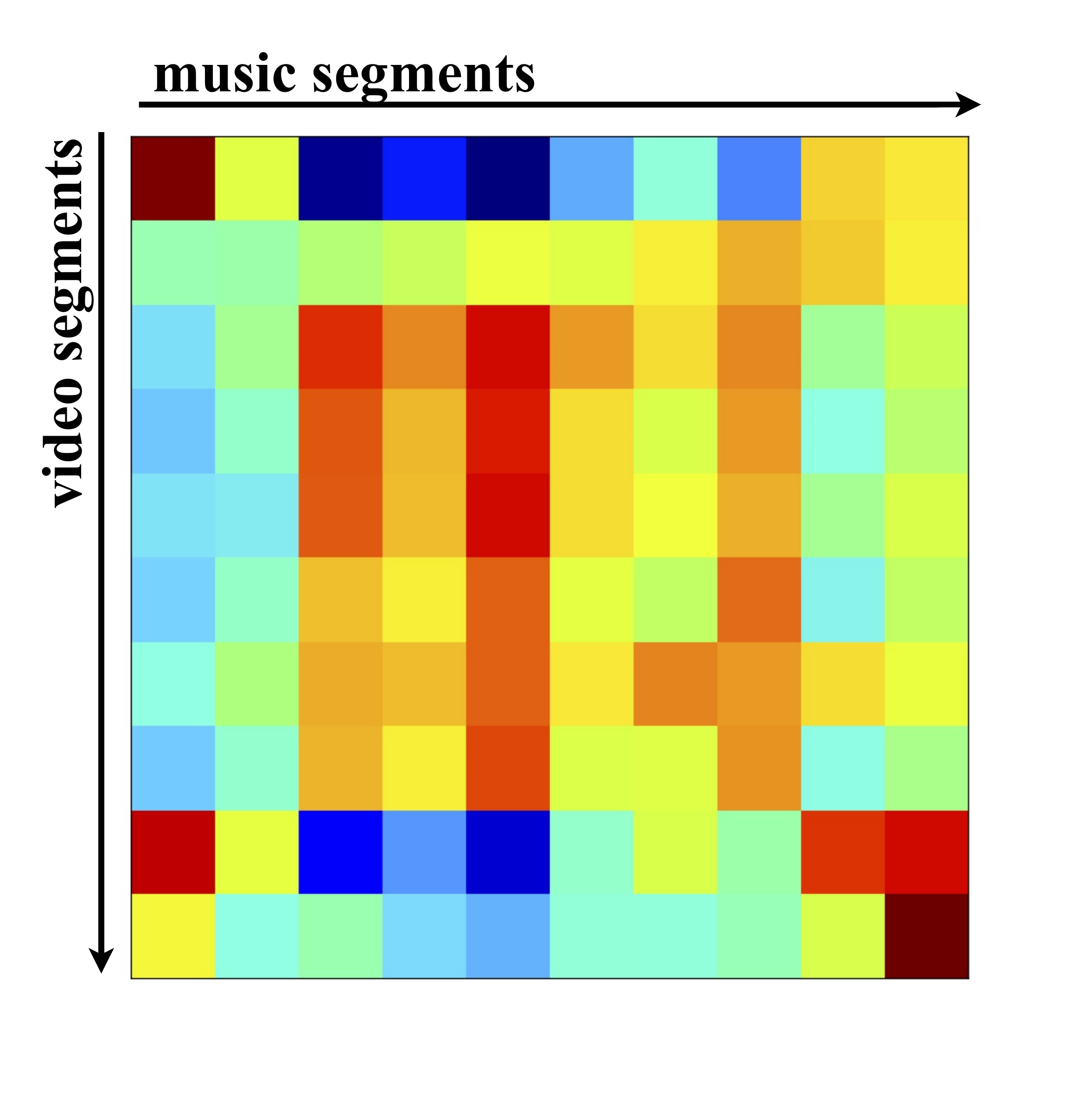}
    \caption{Pairwise similarity matrix $X$ of segment embeddings for the track RnB\textunderscore 11 of the MVD ($S_{OLDA}$ segmentation).
    Warm colors at location $(i,j)$ indicate high similarity between $e^m_{c,i}$ and $e^v_{c,j}$, while cold colors indicate low similarity.}
    \label{fig:ssm}
\end{figure}

\textbf{Validating our segment assumption.}
The assumption that motivated our Seg-VM-Net is that the content of $c^v$ or $c^m$ is not homogeneous over time and therefore can hardly be represented by the timeless embeddings $e^v_c$ or $e^m_c$.
We therefore proposed to represent $c$ as a succession of homogeneous segments $\{c_1, \ldots c_{K_c} \}$ and their respective embeddings.
Assuming that the content is not homogeneous over time is equivalent to say that the similarity between $e^v_{c,k}$ and $e^m_{c,k}$ is higher than between $e^v_{c,k}$ and $e^m_{c,k' \neq k}$.
We illustrate on Figure~\ref{fig:ssm} an example of pairwise similarity matrix $X$ between the music $e^m_{c,*}$ and video segments $e^v_{c,*}$ of a same \ac{av} clip $c$. 
We did not perturb the video, nor the music part of the clip (vanilla scenario).
The horizontal axis represents the sequence of music segments $\{e^m_{c,1}, ..., e^m_{c,K_c} \}$, while the vertical axis represents the sequence of video segments $\{ e^v_{c,1}, ..., e^v_{c,K_c} \}$.
Each coefficient $X_{i,j}$ of the pairwise similarity matrix is computed as the scalar product between $e^m_{c,i}$ and $e^v_{c,j}$.
We display in red the coefficients of high similarity (close to 1) and in blue the ones of low similarity (close to -1).
An ideal system would produce very high values on the diagonal and very low values elsewhere, since only the music and video segments extracted at the same timestamp are associated according to our training criteria.
However, given that the example \ac{av} clip was taken from the test set, we only observe an approximation of this pattern.
Certain segments in the middle of the video seem to be interchangeable, but the very first and last music and video segments are strongly associated.
We assume that the beginning and the end of this \ac{av} clip are very distinctive and play a crucial role in the recommendation.
In this vanilla scenario, we expect the diagonal of $X$ to contain the highest coefficients in case of a match, and therefore we expect $\delta_{trace}$ to be the most appropriate ranking distance (schematized on Figure~\ref{fig:main} (a)).

\textbf{Illustration of the results.}
To better understand the system's performance, we provide in Figure~\ref{fig:demo} some examples of recommendations obtained by our best\footnote{We used $S_{OLDA}$ and $\delta_{trace}$.} system on the MVD (vanilla evaluation scenario).
We give 3 examples of video queries $q^v$ and the suggested music tracks.
We only display the first 3 recommendations provided by the system (from left to right).
As proposed by \cite{Hong2018CBVMR:Constraint}, we select one key frame (picture) of the AV clip to represent both the music track and the video.
On top of each picture, we indicate its Trace distance $\delta_{trace}$ to the query.
Below each picture, we provide the name of the AV clip as provided in the MVD.
To the right of the figure, we display the histogram of the distances between the query and all cross-modal samples.
We highlight in red the histogram bin corresponding to the ground truth sample $q^m$, along with its rank in the list of recommendations.
Note that the exact matching sample of the query was not retrieved in the two first examples, hence the $R@3$ for these examples is zero.
As we see, while the Recall is zero, the recommended music tracks are from a similar music genre as the video query: Pop for the first case, Dubstep for the second which still makes sense in terms of applications.
This shows the limitations of using the Recall to evaluate our systems.
For the last query, the correct music track was retrieved in the first position, hence its $R@3$ is one.

%%%%%%%%%%%%%%%%%%%%%%%%%%%%%%%%%%%%%%%%%%%%%%%%%%%%%%%

\section{Conclusion}
\label{sec:conclusion}

In this study, we proposed a simple, yet efficient Video-to-Music recommendation system, named Seg-VM-Net.
We did so by improving a self-supervised system, the VM-Net.
Inspired by the music supervision use case, we trained our model at the segment level, using those defined by the music and video structure.
At inference time, we then ranked the music tracks according to a segment alignment cost.
With these adaptations, we largely outperformed the clip-level baseline (VM-Net), and the segment clustering aggregation systems.
Additionally, the resulting method makes use of pre-trained features and a lightweight network architecture, thus requiring a highly reasonable training time.
Depending on the complexity of the application scenario, we demonstrated many possible ranking distances, which provide various trade-offs between retrieval time and robustness of the performance.
For example, the Best Trace alignment cost is fast to compute and offers robustness to basic transformations of the query (cropping).
It is therefore the best suited for large catalogs.
The DTW alignment costs render the system robust against other transformations of the queries and targets (stretching), while requiring additional computation time. 
They can be used in scenarios involving medium-size catalogs and little hypotheses and constraints on the allowed media transformations. 
These different options make our system flexible and suitable for applications in music recommendation, music supervision, or automatic video editing.

% Limitations and Perspectives
The main limitation of our method is its coarse temporal resolution, as the music structural segments last, on average, 18.6 seconds.
This does not allow for performing a fine-grained alignment of the two modalities.
Future work will have to include frame-level temporal alignment scores to our ranking distance, as in \cite{Gillet2007OnVideos} (one video frame is usually 40 milliseconds).
The corresponding challenge would require increasing the temporal precision while keeping the training and inference costs reasonable.
Ideally, we should also be able to directly propose an edit of the music tracks to the video query. 
To do so, the recommendation system has to be robust against more diverse transformations of the queries and targets, and to integrate an audio and video salient event detection module.

%%%%%%%%%%%%%%%%%%%%%%%%%%%%%%%%%%%%%%%%%%%%%%%%%%%%%%%
%%%%%%%%%%%%%%%%% BIBLIOGRAPHY %%%%%%%%%%%%%%%%%%%%%%%%
%%%%%%%%%%%%%%%%%%%%%%%%%%%%%%%%%%%%%%%%%%%%%%%%%%%%%%%

\bibliographystyle{IEEEtran}
\bibliography{IEEEabrv, references}

%%%%%%%%%%%%%%%%%%%%%%%%%%%%%%%%%%%%%%%%%%%%%%%%%%%%%%%
%%%%%%%%%%%%%%%%%%%% BIOGRAPHY %%%%%%%%%%%%%%%%%%%%%%%%
%%%%%%%%%%%%%%%%%%%%%%%%%%%%%%%%%%%%%%%%%%%%%%%%%%%%%%%

\newpage

\begin{IEEEbiography}[{\includegraphics[width=1in,height=1.25in,clip,keepaspectratio]{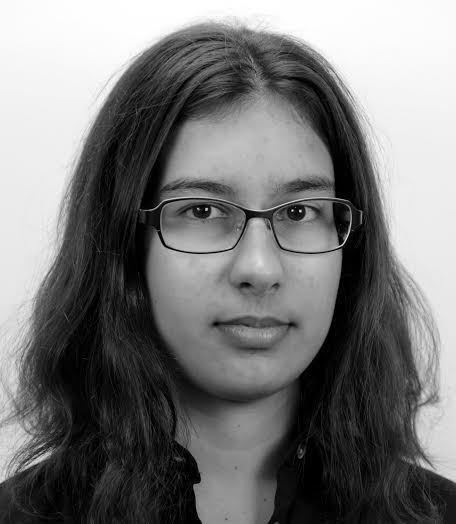}}]{Laure Prétet}
received her Computer Science Engineering degree in  2018 from Télécom Paris, France. She is currently a Ph.D. student at Télécom Paris on the topic of deep learning for music supervision, in collaboration with the start-up company Bridge.audio. Her research interests include music information retrieval, audio processing, multimodal deep learning and self-supervised systems.
\end{IEEEbiography}

\begin{IEEEbiography}[{\includegraphics[width=1in,height=1.25in,clip,keepaspectratio]{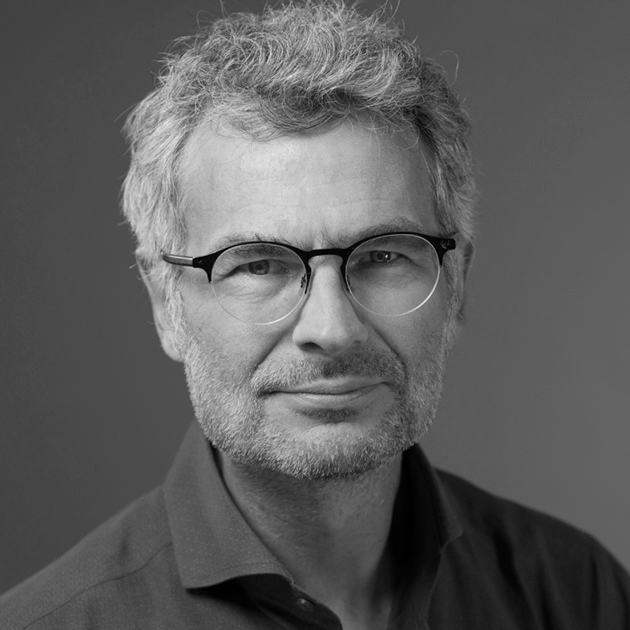}}]{Gaël Richard}
is the executive director of Hi! Paris and Professor at Télécom Paris. 
His research work lies at the core of digitization and is dedicated to the analysis, transformation, understanding and automatic indexing of acoustic signals (including speech, music, environmental sounds) and to a lesser extent of heterogeneous and multimodal signals. In particular, he developed several source separation methods for audio and musical signals based on machine learning approaches.
He has received in 2020 the Grand Prix IMT-Académie des Sciences.
\end{IEEEbiography}

\begin{IEEEbiography}[{\includegraphics[width=1in,height=1.25in,clip,keepaspectratio]{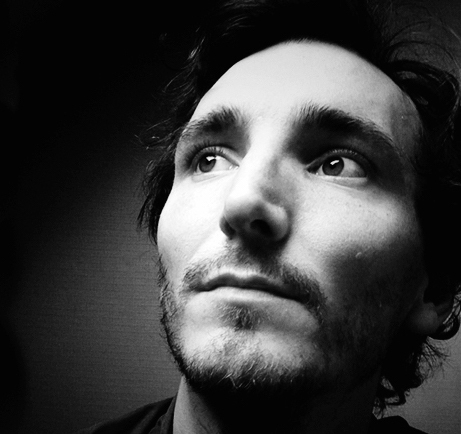}}]{Clément Souchier}
obtained his Master's degree in Entrepreneurship from Université Paris Dauphine in 1998. He is the founder of Soundicate, an early music streaming service, Creaminal, a music supervision agency, and Bridge.audio, a music technology startup. He is a renowned specialist of the music industry with a deep interest in machine learning and music information retrieval.
\end{IEEEbiography}

\begin{IEEEbiography}[{\includegraphics[width=1in,height=1.25in,clip,keepaspectratio]{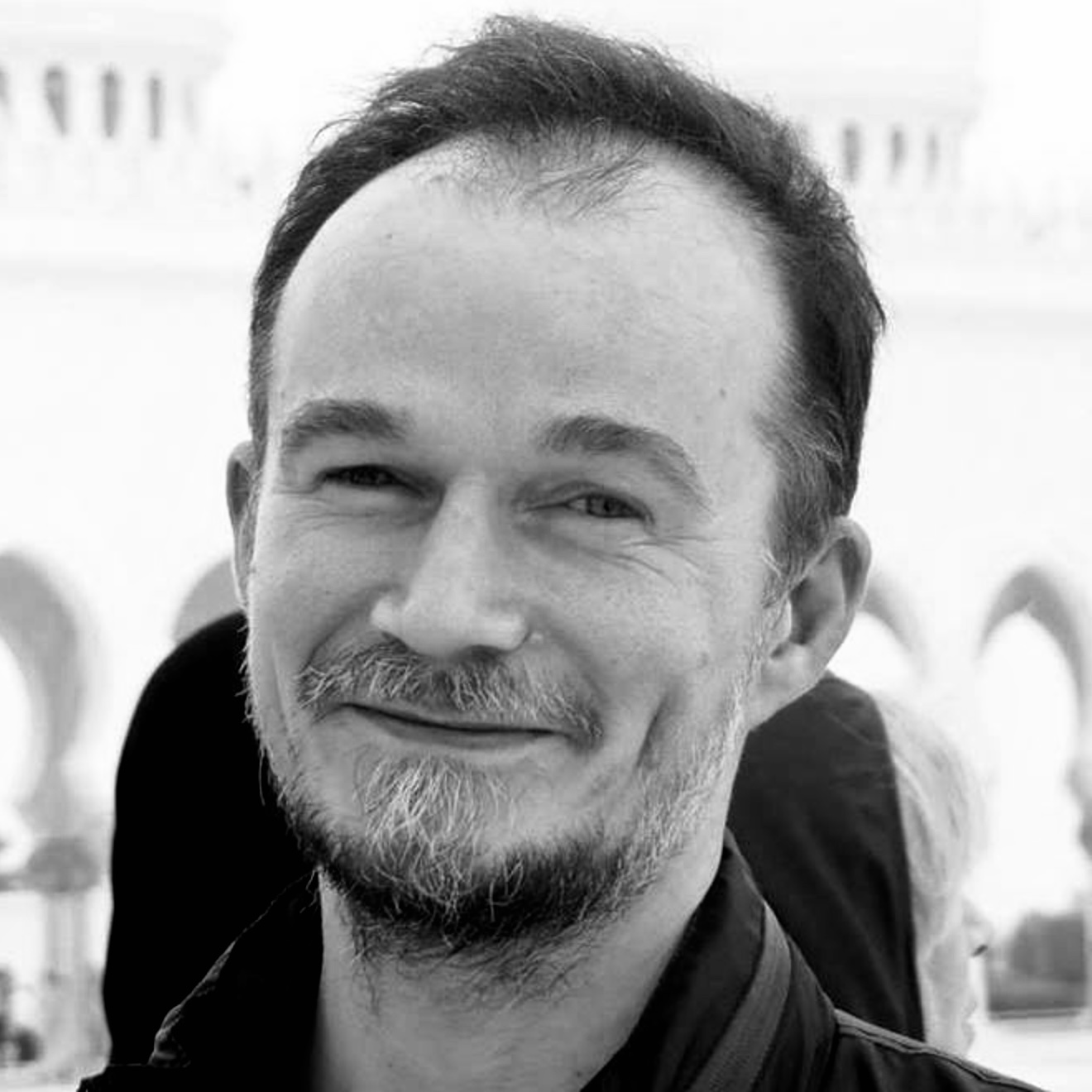}}]{Geoffroy Peeters}
after leading research related to Music-Information-Retrieval at IRCAM from 2001 to 2018, he joined Telecom Paris as a Full Professor.
His main research activities are in the domain of signal processing and machine-learning/deep-learning applied to audio and music. 
He is co-author of the MPEG-7 audio standard and current president of the ISMIR society.
\end{IEEEbiography}

\end{document}